\documentclass[a4paper,11pt]{article}
\pdfoutput=1 

\usepackage{jheppub} 
\usepackage{subfigure}
\usepackage{xcolor}
\usepackage[T1]{fontenc} 

\newcommand{\pT}{\ensuremath{p_{\mathrm{T}}}}
\newcommand{\mT}{\ensuremath{m_{\mathrm{T}}}}
\newcommand{\ET}{\ensuremath{E_{\mathrm{T}}}}
\newcommand{\MET}{\mbox{\ensuremath{\not \!\! \ET}}}
\newcommand{\MEX}{\mbox{\ensuremath{\not \!\! E_{x}}}}
\newcommand{\MEY}{\mbox{\ensuremath{\not \!\! E_{y}}}}

\def\GeV{\ifmmode {\mathrm{\ Ge\kern -0.1em V}}\else \textrm{Ge\kern -0.1em V}\fi}%
\def\GeV{\ifmmode {\mathrm{\ Ge\kern -0.1em V}}\else \textrm{Ge\kern -0.1em V}\fi}%

\title{\boldmath On the Model Dependence of Fiducial Cross Section Measurements in View of Reinterpretations}

\author[a]{Gabriel Facini}
\author[b]{Kyrylo Merkotan}
\author[b]{Matthias Schott\footnote{corresponding author}}
\author[b]{Alexander Sydorenko\footnote{corresponding author}}

\affiliation[a]{University College London, London, UK}
\affiliation[b]{Johannes Gutenberg-University, Mainz, Germany}

\emailAdd{matthias.schott@cern.ch, alexander.sydorenko@cern.ch}

\abstract{
Fiducial production cross sections measurements of Standard Model processes, in principle, provide constraints on new physics scenarios via a comparison of the predicted Standard Model cross section and the observed cross section. This approach received significant attention in recent years, both from direct constraints on specific models and the interpretation of measurements in the view of effective field theories. A generic problem in the reinterpretation of Standard Model measurements are the corrections applied to data to account for detector effects. These corrections inherently assume the Standard Model to be valid, thus implying a model bias of the final result. In this work, we study the size of this bias by studying several new physics models and fiducial phase-space regions. The studies are based on fast detector simulations of a generic multi-purpose detector at the Large Hadron Collider. We conclude that the model bias in the associated reinterpretations is negligible only in specific cases, however, typically on the same level as systematic uncertainties of the available measurements. 
}

\begin{document}

\maketitle


\section{\label{Sec:Intro}Introduction}

With the discovery of the Higgs Boson at the Large Hadron Collider (LHC), we finally have a theory of fundamental particles and their interactions which could be in principle valid up until the Planck Scale. All predictions of the Standard Model (SM) of particle physics have been confirmed in the last decades. Nevertheless, there are many reasons to suggest physics beyond the Standard Model (BSM) i.e., the astrophysical evidence for dark matter or several fine-tuning problems within the theory itself. However, with no evidence of new physics observed at the LHC, it is imperative to consider all potential sources of BSM physics. Several approaches are available: direct searches of new physics signatures; discrepancies in precision measurements of SM observables i.e., couplings, branching-ratios, or particle masses; or systematic probes for deviations from the SM expectation using differential precision measurements of particle production cross sections.

Direct searches for new physics signatures at hadron colliders are typically performed with detector-level, or reconstruction-level, data.  That is, using the calibrated detector response to determine kinematic quantities of particle collision remnants. Relevant kinematic distributions are then compared to the expected SM distributions as well as to the contributions of possible BSM processes. This comparison produces constraints on BSM models when the data agree with the SM predictions. It is important to note that this approach often requires a detailed simulation of the corresponding particle detector to incorporate effects such as experimental resolutions and particle identification efficiencies. The enormous computing resource required for full detector simulations often limits the number of BSM models tested against the collected LHC data. Furthermore, the variables examined are motivated by the BSM models considered and can be suboptimal for other existing models, or models yet to be created. Therefore, a reinterpretation of a direct search in terms of another model ranges from cumbersome to nearly impossible and is highly dependent on additional information made available by the respective collaborations.

An alternative approach to test BSM models based on the comparison of measured cross sections, i.e. observables which are corrected to be independent of detector effects, immediately circumvents the need for detailed detector simulations of BSM models.  The measured cross sections are directly comparable to particle-level predictions before the interaction with the detector. Cross section measurements are typically performed for SM processes, and subsequently used to test theory predictions and tune dedicated Monte Carlo Event generators, e.g. \textsc{Pythia8} \cite{Sjostrand:2007gs}, \textsc{Sherpa} \cite{Gleisberg:2008ta}, \textsc{Herwig} \cite{Bahr:2008pv} or \textsc{MadGraph} \cite{Alwall:2011uj}. The basic idea of a cross section measurement is, in principle, simple and exemplified in the following with the Drell-Yan process $pp\rightarrow Z \rightarrow \mu\mu$ in the muon decay channel. The final state of this process involves two opposite charged muons with a relatively large transverse momentum $\pT$ and a corresponding invariant mass close to the mass of the $Z$ boson $m_Z$. A typical detector-level event selection for this process could require two oppositely charged muons with a minimal $\pT$ of 25 \,GeV and a maximal pseudo-rapidity\footnote{Defined as $\eta = -ln[tan(\theta/3)]$, where $\theta$ is the angle between the particle three-momentum and the positive direction of the beam axis.} of $|\eta|<2.5$ (accounting for the limited detector acceptance) which yield an invariant mass in the range $70<m_{\mu\mu}<110$ \,GeV. This selection defines a fiducial region in a phase-space and can be applied both on detector-level data as well as on particle-level of a MC event generator.  The cross section for any defined fiducial phase-space is given by

\begin{equation}
\label{eqn:cross}
    \sigma _{fid} = \frac{N_{\textrm{Cand}} - N_B}{C\cdot \int L dt},
\end{equation}

where $N_\textrm{Cand}$ is the number of selected signal events in data, $N_B$ is the number of background events, $\int L \text{d}t$ is the integrated luminosity of the corresponding data set, and $C$ is the efficiency correction factor that accounts for the detector response. The latter is estimated with simulated MC samples and defined as the ratio of the expected number of reconstructed events ($N^\textrm{fid}_\textrm{MC-Detector-Level}$) over the number of generated events in the fiducial volume ($N^\textrm{fid}_\textrm{MC-Particle-Level}$),

\begin{equation}
\label{eqn:C}
C = \frac{N^\textrm{fid}_\textrm{MC-Detector-Level}}{N^\textrm{fid}_\textrm{MC-Particle-Level}}.
\end{equation}

The fiducial cross section is related to the inclusive cross section by ${\sigma _{inc} = \sigma _{fid} / A}$, where $A$ is an acceptance correction factor defined as the fraction of generator events that fall into the fiducial volume. The acceptance correction typically includes significant model dependence as one has to extrapolate into a phase-space which is not measured. Hence, to first order, experimental uncertainties affect $C$, while theoretical uncertainties affect $A$. 

The latest measurement of the $Z\rightarrow ll$ boson production cross section, in proton-proton collisions at a center-of-mass energy of 13 TeV, have been performed by the ATLAS and CMS collaborations using slightly different fiducial volumes leading to values of 779$\pm$3(stat.)$\pm$6(sys.) $\pm$16(lumi)\,pb \cite{Aaboud:2016zpd} and 640$\pm$10(stat.)$\pm$20(sys.)$\pm$30(lumi)\,pb \cite{CMS:2015ois}, respectively. Strictly speaking, these measured cross sections are only valid for the neutral Drell-Yan production, since the $C$-factor was derived using the neutral Drell-Yan process. All published cross sections at the LHC exhibit this model dependence as the SM, through MC simulations, is always assumed when deriving $C$-factors. Due to the increased interest in the reinterpretation of published SM cross sections in the view of BSM signatures, questions concerning the impact of model dependence become more and more important. It should be noted that we limit our discussion to simple fiducial measurements, however, the reinterpretation of (unfolded) differential cross-section measurements will be subject to similar, if not larger, model dependencies.

Discussions on the challenges and limitations of SM cross-section reinterpretations in the view of new physics are ongoing since several years within the community and several discussion workshops have been organized. In this article, we summarize and exemplify the main arguments and quantify for the first time the model dependence using more than twenty SM and BSM processes, ranging from supersymmetric scenarios, to leptoquarks, to the impact of selected 6-dimensional effective field theory operators in more than ten fiducial volumes. In Section \ref{Sec:Samples}, benchmark physics models, as well as the detector simulation and associated uncertainties, are introduced. The fiducial volumes under study typically target SM processes or potential signal regions of BSM models and are presented in Section \ref{Sec:PhaseSpace}. Section \ref{Sec:Results} discusses the impact of expected experimental uncertainties and correction factor model dependence for cross section measurements regarding BSM physics reinterpretations.


\section{\label{Sec:Samples}Simulated Data Samples and New Physics Models}

\subsection{Physics Models}

In order to study the model dependence of $C$-factors for different fiducial volumes, several different SM and BSM processes in proton-proton collisions at a center-of-mass energy of 13 TeV were simulated. The Drell-Yan $W$ and $Z$ boson production and diboson production of $WW$ and $WZ$  were produced in the electron and muon decay channels, as well as the production of top-quark pairs $t\bar t$ in the fully leptonic ($t\bar t \rightarrow b\bar b l^+l^-\nu\nu$) and semi-leptonic ($t\bar t \rightarrow b\bar b q \bar q l\nu$) decay channels. These processes were simulated using the \textsc{MadGraph5} \cite{Alwall:2011uj} and the \textsc{Pythia8} \cite{Sjostrand:2007gs} MC event generator, the CT10nlo PDF set \cite{Dulat:2015mca} (NNPDF2.3 \cite{Ball:2017nwa} for pure Pythia) and the standard \textsc{Pythia8} parton shower tunes. 

In the following, processes containing at least one lepton, defined as a muon or electron $l^\pm=e^\pm, \mu^\pm$, are considered. The decays of $\tau$ leptons have not been included.

In addition to SM processes, a variety of BSM models, including additional dimensional-6 effective field theory (EFT) operators, were simulated using either the \textsc{Pythia8} or the \textsc{MadGraph5} event generator. Since BSM scenarios typically involve several model parameters, e.g. mass- or mixing- parameters of hypothetical new particles, several benchmark points in each BSM scenario were studied.

One of the most prominent BSM models are inspired by GUT theories \cite{Ellis:1981tv, deBoer:1994dg} and predict the existence of leptoquarks (LQs) \cite{Schrempp:1984nj, Pati:1974yy, Dimopoulos:1979es, Dorsner:2016wpm}. LQs are new elementary particles that decay into one lepton and one quark. A continuous mixing parameter $\beta$ controls the lepton flavor in the decay where $\beta=1$ yields charged lepton decays and $\beta=0$ gives decays to neutrinos only. LQs are produced either in pairs via the strong interaction or singly via an electroweak coupling. First and second generation LQ pair-production with masses between 0.4 and 2.0\,TeV and a mixing parameter $\beta=1$, leading to di-lepton (electron or muon) and di-jet final states, has been studied.  

A fourth generation (4G) of heavy fermions \cite{Djouadi:2012ae, Kribs:2007nz, Frampton:1999xi, Martin:2009bg} would have a significant impact on the electroweak symmetry breaking and substantial CP violation in the 4x4 CKM matrix playing a crucial role in understanding the baryon asymmetry in the universe. Hence several searches for fourth generation fermions have been conducted and are still ongoing. Pair-production of heavy up-type quarks $t'$ with masses of 200, 400, 600 and 800\,GeV, decaying via $t'\rightarrow Wb$ leading to an overall final state of $t't'\rightarrow WbWb \rightarrow bb f\bar{f}'  l \nu_l$ in the lepton + jets channel was studied.

Several extensions of the SM predict new heavy gauge bosons ($W'$ and $Z'$) with significantly higher masses than the $W$ and $Z$ boson, i.e., models with extra dimensionsg. \cite{ArkaniHamed:1998rs, Randall:1999ee, Altarelli:1989ff, London:1986dk, delAguila:2010mx}. Searches for these particles are a cornerstone of the search programs at collider experiments. New gauge bosons $W'$ and $Z'$ with masses of 0.5, 1.0, 1.5 and 2.0 TeV in the leptonic decay channels $W'\rightarrow l\nu$ and $Z'\rightarrow l^+l^-$, respectively, are discussed in this article.

Even though no direct signs of supersymmetric (SUSY) particles could be found so far\cite{Aad:2014wea, Chatrchyan:2013iqa}, the corresponding models \cite{Fairbairn:2006gg, Djouadi:1998di, Aitchison:2005cf} are extremely popular due to their intrinsic ability to solve fine-tuning problems of the SM as well as provide candidates for the observed dark matter content of the universe. Since it is impossible to study all possible final states of supersymmetric scenarios because of the huge model-parameter space, studies here are focused in an MSSM scenario on the production of top squarks $\tilde{t}\tilde{t'}$ and their subsequent decay to top quarks and neutralinos $\tilde{t} \tilde{t'} \rightarrow t \tilde{\chi}^{0}_1 \bar{t} \tilde{\chi}^0_1 \rightarrow b\bar{b} WW \tilde{\chi}^0_1 \tilde{\chi}^0_1 \rightarrow \tilde{\chi}^0_1 \tilde{\chi}^0_1 b\bar{b} l^+ l^- \nu_l \bar{\nu_l}$. 

If the actual energy scale of BSM processes is beyond the reach of the LHC such that direct production is not possible, effective field theories (EFTs) parameterize the BSM impact on observables. The EFT approach to categories and interpret typical SM final states received significant attention in recent years \cite{Burgess:2007pt, Brivio:2017vri}. The impact of EFT parameter variations on SM signatures is of particular interest since several groups are already using published measurements to constrain EFT parameters \cite{Berthier:2015gja}, where these measurements have been performed by assuming the SM as underlying theory. In this work, we study the impact of the $\mbox{Tr}[W_{\mu\nu}W^{\nu\rho}W_{\rho}^{\mu}]$ and $(D_\mu\Phi)^\dagger W^{\mu\nu}(D_\nu\Phi)$ operators in the diboson WW and WZ final states. Table \ref{tab:ModelOverview} contains a summary of all simulated processes and decay channels.

\begin{table}[h]
\center
\footnotesize
\caption{Overview of generated samples and processes used in this study. The lepton decay $l$ refers exclusively to $(l=e,\mu)$}
{
 \begin{tabular}{l | c|c|c|c}
\hline
Sample Name			& Decay-Chain								& (Model) Parameter(s)	& O($\alpha_s$)	&	Generator				\\
\hline
Drell-Yan $Z/\gamma*$	& $Z/\gamma*\rightarrow l^+ l^-$				& $60<m_{ll}<110$\,GeV,	& NLO			& \textsc{MadGraph+Pythia}	\\
					&										& 100, 200, 500\,GeV$<m_{ll}$		&				& 						\\
$W^\pm$				& $W^\pm \rightarrow l^\pm \nu$				& -					& LO/NLO			& \textsc{MadGraph+Pythia}	\\
$t\bar t$ (di-lep.)		& $t\bar t \rightarrow l^+ \nu b l^- \nu \bar{b}$		& -					& LO/NLO			& \textsc{MadGraph+Pythia}	\\
$t\bar t$ (semi.-lep.)		& $t\bar t \rightarrow l^+ \nu b q \bar{q}' \bar{b}$	& -					& LO/NLO			& \textsc{MadGraph+Pythia}	\\
$WW$ (di-lep.)			& $W^+W^-\rightarrow l^+ \nu l^- \nu$				& -					& LO/NLO			& \textsc{MadGraph+Pythia}	\\
$WW$ (semi.-lep.)		& $W^+W^-\rightarrow l^\pm \nu q \bar{q}'$				& -					& LO/NLO			& \textsc{MadGraph+Pythia}	\\
$WZ$ (di-lep.)			& $W^\pm Z\rightarrow l^\pm \nu l^+ l^-$				& -					& LO/NLO			& \textsc{MadGraph+Pythia}	\\
$WZ$ (semi.-lep.)		& $W^\pm Z\rightarrow l^\pm \nu q \bar{q}$				& -					& LO/NLO			& \textsc{MadGraph+Pythia}	\\
\hline
$WW$ (EFT-1)			& $WW\rightarrow l^+ \nu l^- \nu$				& 	$c_{WWW}/\Lambda^2$ = -35	& LO				& \textsc{MadGraph+Pythia}	\\
$WW$ (EFT-2)			& $WW\rightarrow l^+ \nu l^- \nu$				& 	$c_{W}/\Lambda^2$ = 40		& LO				& \textsc{MadGraph+Pythia}	\\
$WZ$ (EFT-1)			& $W^\pm Z\rightarrow l^\pm \nu l^+ l^-$				& 	$c_{WWW}/\Lambda^2$ = -35	& LO				& \textsc{MadGraph+Pythia}	\\
$WZ$ (EFT-2)			& $W^\pm Z\rightarrow l^\pm \nu l^+ l^-$				& 	$c_{W}/\Lambda^2$ = 40		& LO				& \textsc{MadGraph+Pythia}	\\
\hline
$Z'$					& $Z'\rightarrow l^+ l^-$						& $m_{Z'}=0.5, 1.0, 1.5$ TeV			& LO			& \textsc{Pythia}	\\
$W'$					& $W'^\pm \rightarrow l^\pm \nu$				& $m_{W'}=1.0,1.5,2.0$ TeV			& LO			& \textsc{Pythia}	\\
$4^{th}$-Gen. Quark		& $\bar{t'} t' \rightarrow b \bar{b} f \bar{f}' l^- \bar{\nu}_l $	& $m_{q4}=0.2, 0.4, $		& LO			& \textsc{MadGraph+Pythia}	\\
					&										&	$0.6, 0.8$ TeV			&				&			\\
LQ ($1^{st}$-Gen)     	&  $LQ \bar{LQ}\rightarrow e^+ u e^- \bar{u}$ 			& $m_{LQ}=0.4, 0.6,$  			& LO 		& \textsc{Pythia} \\
					&										&	$1.0, 2.0$ TeV				&				&			\\
LQ ($2^{nd}$-Gen)		& $LQ \bar{LQ}\rightarrow \mu^+ c \mu^- \bar{c}$	& $m_{LQ}=0.5, 1.0,$ 			& LO			& \textsc{Pythia}	\\
					&										&	$1.5, 2.0$ TeV			&				&			\\
SUSY				& $\tilde{t} \tilde{t'}  \rightarrow \tilde{\chi}^0_1 \tilde{\chi}^0_1 +$		& \textsc{MSSM SLHA2}					& LO			& \textsc{MadGraph+Pythia}	\\
					& $+b\bar{b} l^+ l^- \nu_l \bar{\nu_l}$				& 					& 			& 	\\
\hline
\end{tabular}

\label{tab:ModelOverview} 
}
\end{table}

\subsection{Detector Simulation and Uncertainties}

The detector response was simulated using the \textsc{Delphes} \cite{deFavereau:2013fsa} framework and all the nominal ATLAS detector simulation settings except for the lepton isolation requirements\footnote{The presented results have been also cross-checked for several Standard Model signal processes simulated using the full Geant4 simulation of a different LHC experiment, available thanks to the open data project \cite{CERN:OpenData}}. Instead, one loose and one tight customized lepton isolation criteria were defined. Tight isolation is satisfied if the \pT -sum of charged particles within $\Delta R<0.2$ around the signal lepton divided by the lepton $\pT$ is smaller than $0.2$. Loose isolation requires a value smaller than $0.3$. 

In order to approximate the experimental uncertainties on the derived $C$-factors for the different samples, additional uncertainties are assumed for the lepton- and b-tag efficiencies, as well as the energy scales of electrons, muons, jets and the missing transverse momentum observable $\MET$. The latter is a measure of transverse momenta of particles that leave the detector undetected (e.g. neutrinos) and is defined as the negative vector sum of the transverse momentum of all identified particles in the event.

The uncertainty values used were motivated by SM measurements \cite{Aaboud:2016btc, Aad:2010ey, Aad:2012qf} and are summarized in Table \ref{tab:ExpUnc}. They certainly do not give a complete estimation of the true experimental uncertainties, but rather relay the order of magnitude of the expected effects. All uncertainties have been applied on object-level and taken uncorrelated among each other. Basic kinematic dependencies of the assumed uncertainties have been taken into account.

\begin{table}[h]
\center
\footnotesize
\caption{Overview of detector related uncertainties considered in this study.}
{
\begin{tabular}{l | c|l|c}

\hline
Quantity					& Relative eff.				& Quantity						& Relative scale			\\
						& uncertainty				& 							& uncertainty				\\
\hline
Electron/Photon eff.		& 0.5\%					& Electron/Photon energy scale	& 0.1\%					\\
Muon efficiency				& 0.5\%					& Muon momentum scale			& 0.1\%					\\
Lepton isolation eff.		& 0.3\%					& Jet energy scale				& 4\% (for $E_\text{T} <$40\,GeV)	\\
b-tagging efficiency			& 4.0\%					& 							& 2\% (for $E_\text{T}>$40\,GeV)	\\
						& 						& $\MET$  scale				& 4\% (for $E_\text{T}<$40\,GeV)	\\
						& 						& 							& 2\% (for $E_\text{T}>$40,\GeV)	\\
\hline
\end{tabular}
\label{tab:ExpUnc}
}
\end{table}


\section{\label{Sec:PhaseSpace}Signal Selection and Fiducial phase-space Regions}

It is impossible to study the model dependence of the $C$-factors used in Equation \ref{eqn:cross} for all possible final states and scenarios.  The model dependence of the $C$-factors was therefore studied with eight selected fiducial phase-space regions dedicated to SM processes, five fiducial phase-space regions aiming for direct searches of new elementary particles as well as four differential distributions typically used to constrain EFT parameters. The fiducial phase-space definitions used in this study are summarized in Table \ref{tab:FidDefinitions}. The selected phase-space regions were chosen to cover a large variety of final states with a range of final state objects and multiplicities, as well as in different kinematic regimes. Therefore, general conclusions can be drawn from the corresponding studies.

The same kinematic requirements are applied at particle-level and detector-level. All jets are reconstructed using the anti-$k_\text{T}$ algorithm \cite{Cacciari:2008gp} with a radius parameter $R$ of $0.4$. The jets are required to be within a rapidity of $|y|<4.0$ with a minimal $\pT$ of 30\,GeV. The basic selection requirements for leptons is a minimal $p_\text{T}>25$\,GeV within a pseudo-rapidity value of $|\eta|<2.4$. In addition to the kinematic lepton selection, the tight lepton isolation requirements are applied for reconstructed leptons. To not double-count objects, overlap removal is applied on particle-level and detector-level objects, discarding any jet that is closer than $\Delta R =0.4$ to a lepton. The transverse mass, $m_\text{T}$, in events with significant \MET\,is defined as

\begin{equation}
\label{eqn:mt}
m_\text{T} = \sqrt{(\MET+\sum_i p_\text{T}(l_i))^2-(\MEX+\sum_i p_x(l_i))^2-(\MEY+\sum_i p_y(l_i))^2},
\end{equation}
where $l_i$ denote signal leptons in the event. Selection requirements on the number of leptons and jets are always exclusive, i.e., events with three leptons in the fiducial region are discarded in a selection that requires (exactly) two leptons.
 
\begin{table}[h]
\center
\footnotesize
\caption{Overview of the fiducial phase-space regions used in this study aiming for different signal selections. The kinematics variables used follow the standard definitions: transverse mass is defined in eq. \ref{eqn:mt}; $m_{ll}$ and $p_\text{T}(ll)$ describes the invariant mass and the invariant transverse momentum $p_\text{T}$ of two signal leptons in an event; $n_l$, $n_{jet}$ and $n_{b-jet}$ are the number of leptons, jets and identified b-jets per event, respectively; the observable $S_\text{T}$ is defined as the scalar sum of all selected jet and lepton transverse energies in the event.}
{
    \begin{tabular}{l | c}
\hline
\multicolumn{2}{c}{\bf Standard Model regions}																			\\	
\hline
Scenario/Process				& Fiducial phase-space definitions														\\
\hline
$Z/\gamma^*$					& $n_l=2$, $p_\text{T}(l)>25$\,GeV, $|\eta(l)|<2.4$, $70< m_{ll}<110$\,GeV								\\
\hline
$W^\pm$						& $n_l=1$, $p_\text{T}(l)>25$\,GeV, $|\eta(l)|<2.4$, \MET$>30$\,GeV, $m_\text{T}>40$\,GeV				\\
\hline
$W^+W^-$ (di lep.)					& $n_l=2$, $p_\text{T}(l)>25$\,GeV, $|\eta(l)|<2.4$, \MET $>30$\,GeV, 			\\
							& $p_\text{T}(ll)>30$\,GeV,	$m_{ll}-m_Z >30$\,GeV			\\
\hline
$W^\pm Z$ (di lep.)					& $n_l=3$, $p_\text{T}(l)>25$\,GeV, $|\eta(l)|<2.4$, \MET $>25$\,GeV, $m_\text{T}(W)>30$\,GeV				\\
\hline
$t\bar t$ (semi. lep.)				& $n_l=1$, $p_\text{T}(l)>25$\,GeV, $|\eta(l)|<2.4$, \MET $>40$\,GeV 				\\
							& $n_{b-\textrm{jet}}\geq1$, $n_{\textrm{jet}}\geq3$, $p_\text{T}(\textrm{jet})>30$\,GeV, $|\eta(\textrm{jet})|<4.0$							\\
\hline
$t\bar t$ (di lep.)				& $n_l=2$, $p_\text{T}(l)>25$\,GeV, $|\eta(l)|<2.4$, \MET $>60$\,GeV, $|m_{ll}-m_Z| >30$\,GeV, 			\\
							& $m_{ll}>10$\,GeV, $n_{b-\textrm{jet}}\geq1$, $n_{\textrm{jet}}\geq2$, $p_\text{T}(\textrm{jet})>30$\,GeV, $|\eta(\textrm{jet})|<4.0$							\\
\hline
\hline
\multicolumn{2}{c}{\bf BSM search regions}																			\\	
\hline
$Z'$ 							& $n_l=2$, $p_\text{T}(l)>25$\,GeV, $|\eta(l)|<2.4$, $m_{ll}>200$\,GeV								\\
\hline
        ${W'}^\pm$ 							& $n_l=1$, $p_\text{T}(l)>25$\,GeV, $|\eta(l)|<2.4$, \MET $>60$\,GeV, $m_\text{T}>500$\,GeV				\\
\hline
LQ 							& $n_l=2$, $p_\text{T}(l)>25$\,GeV, $|\eta(l)|<2.4$, \MET $>25$\,GeV, $S_\text{T}>400$\,GeV				\\
							& $n_{\textrm{jet}}\geq2$, $p_\text{T}(\textrm{jet})>30$\,GeV, $|\eta(\textrm{jet})|<4.0$, $m_{l,\textrm{jet}}>300$\,GeV						\\
\hline
4th Generation 					& $n_l=1$, $p_\text{T}(l)>25$\,GeV, $|\eta(l)|<2.4$, \MET $>35$\,GeV, $ \MET + m_\text{T}>60$\,GeV		\\
							& $n_{b-\textrm{jet}}\geq1$, $n_{\textrm{jet}}\geq3$, $p_\text{T}(\textrm{jet})>30$\,GeV, $|\eta(\textrm{jet})|<4.0$,	\\
                                                        & $p_\text{T}(\textrm{jet}^\textrm{lead})>60$\,GeV, $S_\text{T}>400$\,GeV		\\
\hline
SUSY						& $n_l=2$, $p_\text{T}(l)>25$\,GeV, $|\eta(l)|<2.4$												\\
							&$p_\text{T}(\textrm{jet})>30$\,GeV, $|\eta(\textrm{jet})|<4.0$, $n_{\textrm{jet}}\geq2$, \MET $>150$\, GeV							\\
\hline
\hline
\multicolumn{2}{c}{\bf EFT sensitive regions}																			\\	
\hline
        $W^+W^-$ (EFT-1 Sel.)				& Standard $W^+W^-$ (di lep.) + $p_\text{T}(\textrm{lep}^\textrm{lead})>100$\,GeV										 \\
\hline
$W^+W^-$ (EFT-2 Sel.)				& Standard $W^+W^-$ (di lep.) + $m_\text{T}(WW)>200$\,GeV											 \\
\hline
$W^\pm Z$ (EFT-1 Sel.)				& Standard $W^\pm Z$ (di lep.) + $p_\text{T}(\textrm{lep}^\textrm{lead})>80$\,GeV										 \\
\hline
$W^\pm Z$ (EFT-2 Sel.)				& Standard $W^\pm Z$ (di lep.) + $m_\text{T}(WZ)>250$\,GeV											 \\
\hline
\end{tabular}
\label{tab:FidDefinitions}
}
\end{table}


\section{\label{Sec:Results}Model Dependencies}

The possibility to reinterpret a measured fiducial cross section as a BSM physics exclusion limit depends mainly on the similarity of the $C$-factors of the process assumed to perform the measurement and the $C$-factor of the BSM process. For example, 800 observed events for the SM process X in a 100 pb$^{-1}$ data set, and $C$-factor of $C_X=0.8$, leads to a measured fiducial cross section of ${\sigma_X = 800/(0.8\cdot100)=10\,}$pb. Assuming a predicted cross section of 8\,pb for process X in the SM, the measurement can be used to constrain BSM scenarios Y (with a $C$-factor of $C_Y$) which would enhance the measured cross section of events in the fiducial region. In the example expressed, the difference between the expected and observed cross sections of $2\,pb$ limits the cross section of model Y \footnote{Of course, uncertainties must be properly accounted for but are left out of the example for simplicity.}. If the correction factors $C_X$ and $C_Y$ are similar, then the measured cross section can be directly used to place a limit on model Y. However, if the detector correction factor differs largely from the SM expectation, i.e., $C_Y=0.4$, the reinterpretation will lead to a false conclusion on the validity of model Y by a factor of 2 in the above example. It should be noted that a reinterpretation for a given process is perfectly fine, if the $C$-factor for this process is known. Since these $C$-factors are generally not known, it is often assumed that the $C$-factors for different processes are similar. In this work, we probe to which extent this assumption holds, i.e. we study how the detector correction factors for different processes in a given signal selection differ and draw general conclusions. 

\subsection{Standard Model Processes \label{sec:ResultSM}}

First, $C$-factors for different SM processes in phase-space regions typically used in measurements are studied. The measurement of the $Z$ boson cross section, defined by the fiducial volume of Table \ref{tab:FidDefinitions}, is an example to illustrate several common aspects which also hold generally true. The signal process implies two leptons in the fiducial region - both on particle level as well as detector level. The leptonic decay channel in top-quark pair production, as well as leptonic decays in the $WW$, $WZ$ and $ZZ$ diboson production, have to be considered as potential processes that contribute events to the fiducial region on both particle- and detector-level. The derived $C$-factors for the $Z$ boson signal and the background processes are summarized in Table \ref{tab:CFactorsForZ} for both loose and tight lepton isolation requirements. Firstly, we observe significantly larger $C$-factors for the $WZ$ and $ZZ$ production as these processes have more than two leptons in the final state. At particle-level, events with three or four leptons can enter the fiducial volume when one or two leptons are outside the fiducial lepton definition. At detector-level, events with three or four leptons in the fiducial region at the particle-level are counted in the selection when only two leptons are reconstructed. Since there is no requirement on the connection between particle- and detector-level on an event-by-event basis for $C$-factors, there is an overall increase of the corresponding $C$-factors when the lepton multiplicity of the process in question is larger than the fiducal region definition. A first conclusion is drawn: one ought only reinterpret a measurement in terms of BSM processes which have the same final state objects multiplicity as the SM process. In particular, this is important for final state objects that have an associated reconstruction efficiency that differs from unity, i.e., the number of leptons, photons, and heavy-flavor jets. Thus in all further studies, we explicitly require events to have the same number of inclusive truth leptons as the signal region of interest.

The second observation in Table \ref{tab:CFactorsForZ} concerns the isolation requirements. Processes with much hadronic activity in the final state, such as the decay of top-quark pairs, tend to lead to less isolated leptons in the final state compared to final states with less hadronic activity. Hence the $C$-factors for the $Z/\gamma*$ and $WW$ processes are more similar to one another than for $t\bar t$, in particular when requiring tight isolation. Differences from the isolation requirements effect are generally less pronounced when only loose lepton isolation is required. Hence, the amount of hadronic activity, e.g. the number and the energies of particle jets in the given process, should always be considered if a direct reinterpretation is performed. One possible solution to overcome this model dependence is to use very tight isolation requirements and include those in the fiducial volume definition.

\begin{table}[h]
\center
\footnotesize
\caption{Derived $C$-factor including statistical uncertainties for the fiducial volume of a typical Z boson cross section measurement for various SM processes. The experimental uncertainties are expected to be highly correlated. Note that the particle-level requirement on the number of leptons is not applied.}
{
    \begin{tabular}{l | c | c | l | c | c}
\hline
Process							&	$C$-factor	& $C$-factor 	& Process		&	$C$-factor	& $C$-factor \\
								&	(tight iso.)	& (loose iso.)	& 			&	(tight iso.)	& (loose iso.)\\
\hline
$Z/\gamma*\rightarrow \mu^+\mu^-$		&	$0.826\pm 0.001$		& $0.827\pm 0.001$	& $Z/\gamma*\rightarrow e^+e^-$		&	$0.696\pm 0.002$		& $0.697\pm 0.002$	\\
\hline
$t\bar t\rightarrow\mu^+\nu b\mu^-\nu b$	&	$0.826\pm 0.005$		& $0.834\pm 0.005$	& $t\bar t\rightarrow e^+\nu be^-\nu b$	&	$0.708\pm 0.007$		& $0.715\pm 0.007$ \\
$W^+W^- \rightarrow \mu^+\nu\mu^-\nu$		&	$0.825\pm 0.017$		& $0.826\pm 0.017$	& $W^+W^- \rightarrow e^+\nu e^-\nu$		&	$0.686\pm 0.022$		& $0.687\pm 0.022$	\\
$W^\pm Z\rightarrow \mu^\pm\nu\mu^+\mu^-$	&	$1.037\pm 0.012$		& $1.033\pm 0.012$	& $W^\pm Z\rightarrow e^\pm\nu e^+e^-$		&	$0.872\pm 0.09$		& $0.874\pm 0.09$	\\
$ZZ\rightarrow \mu^+\mu^-\mu^+\mu^-$	&	$1.129\pm 0.032$		& $1.131\pm 0.032$	& $ZZ\rightarrow e^+e^-e^+e^-$		&	$1.232\pm 0.051$		& $1.235\pm 0.051$	\\
\hline
\end{tabular}

\label{tab:CFactorsForZ}
}
\end{table}

\begin{figure}[htb]
\begin{center}
\includegraphics[width=7.3cm]{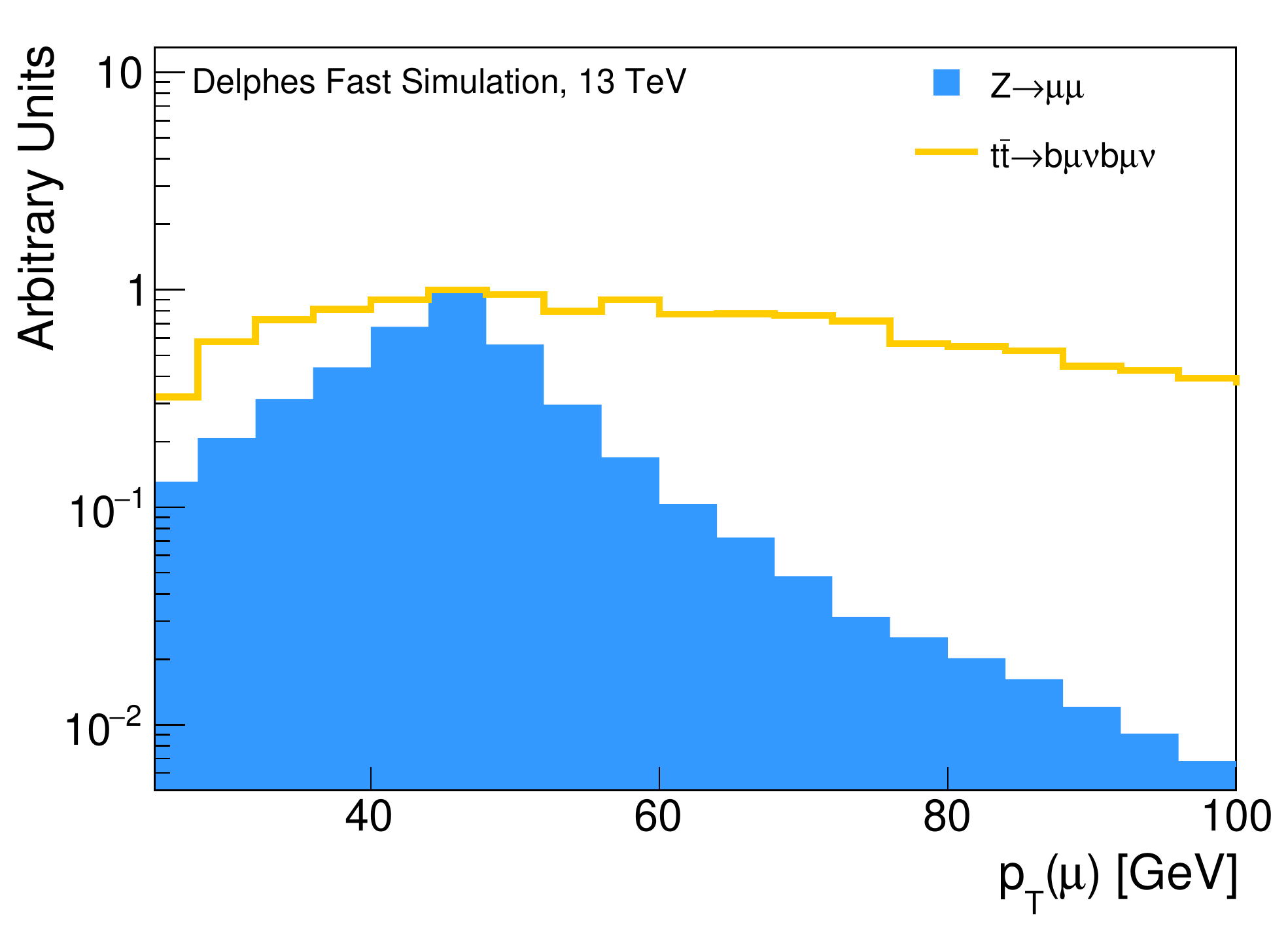} \hspace{0.1cm}
\includegraphics[width=7.3cm]{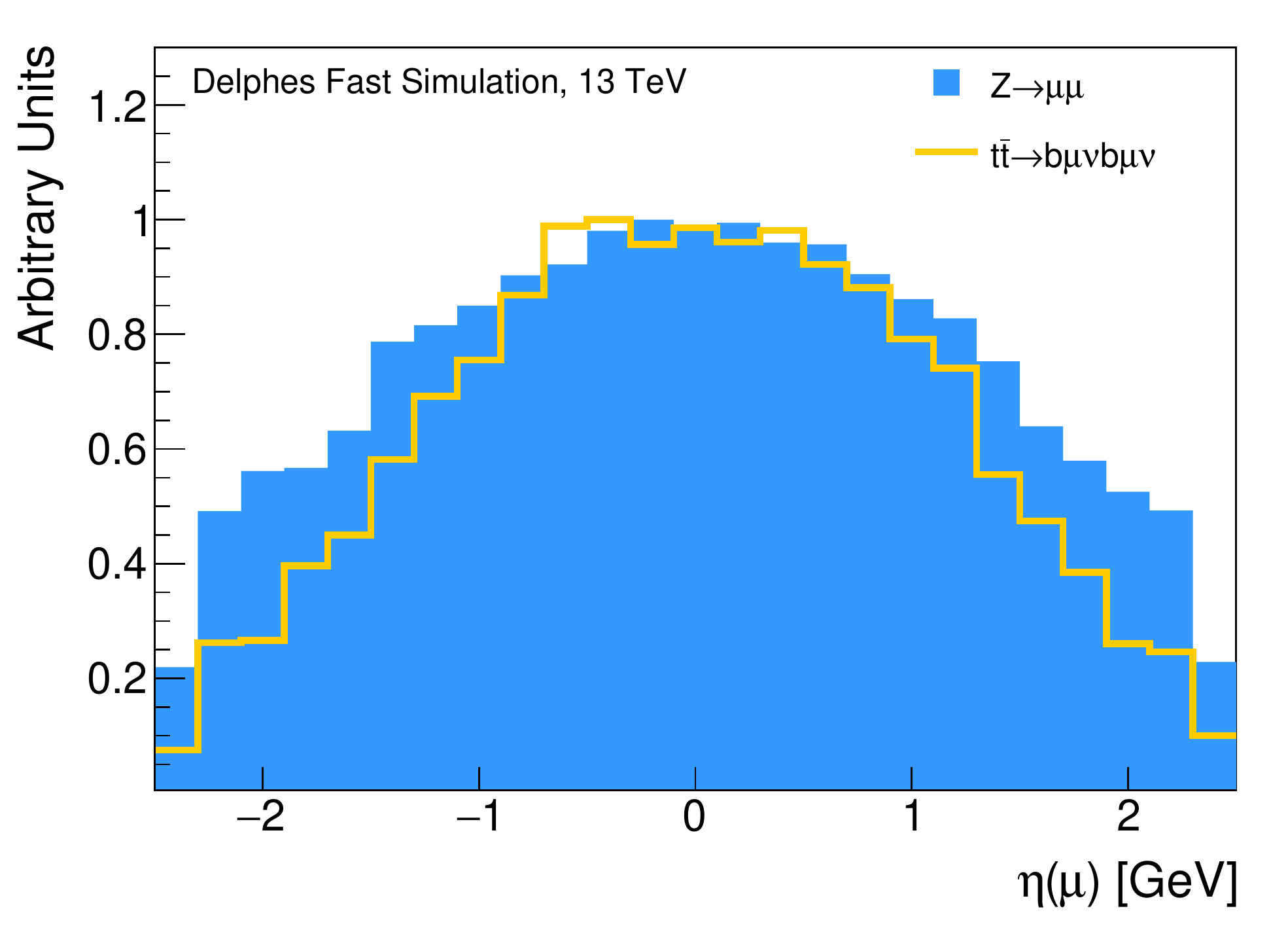}
\caption{\label{fig:KinDistributions} Normalized distribution of reconstructed decay leptons of Z bosons and $t\bar t$ processes for $\pT$ (left) and $\eta$ (right).}
\end{center}
\end{figure}

The remaining differences of the $C$-factors presented in Table  \ref{tab:CFactorsForZ} are due to kinematic differences of the decay leptons, illustrated in Figure \ref{fig:KinDistributions}. These (different) distributions are convoluted with the relevant detector $\eta$ and $\pT$ dependent efficiencies and yield differences in the $C$-factors. Typically, there is only a small $\pT$ dependence for lepton reconstruction efficiencies, and given the similar $\eta$ distributions, the resulting differences on $C$ are expected to be moderate. 

\begin{figure}[htb]
\begin{center}
\includegraphics[width=7.3cm]{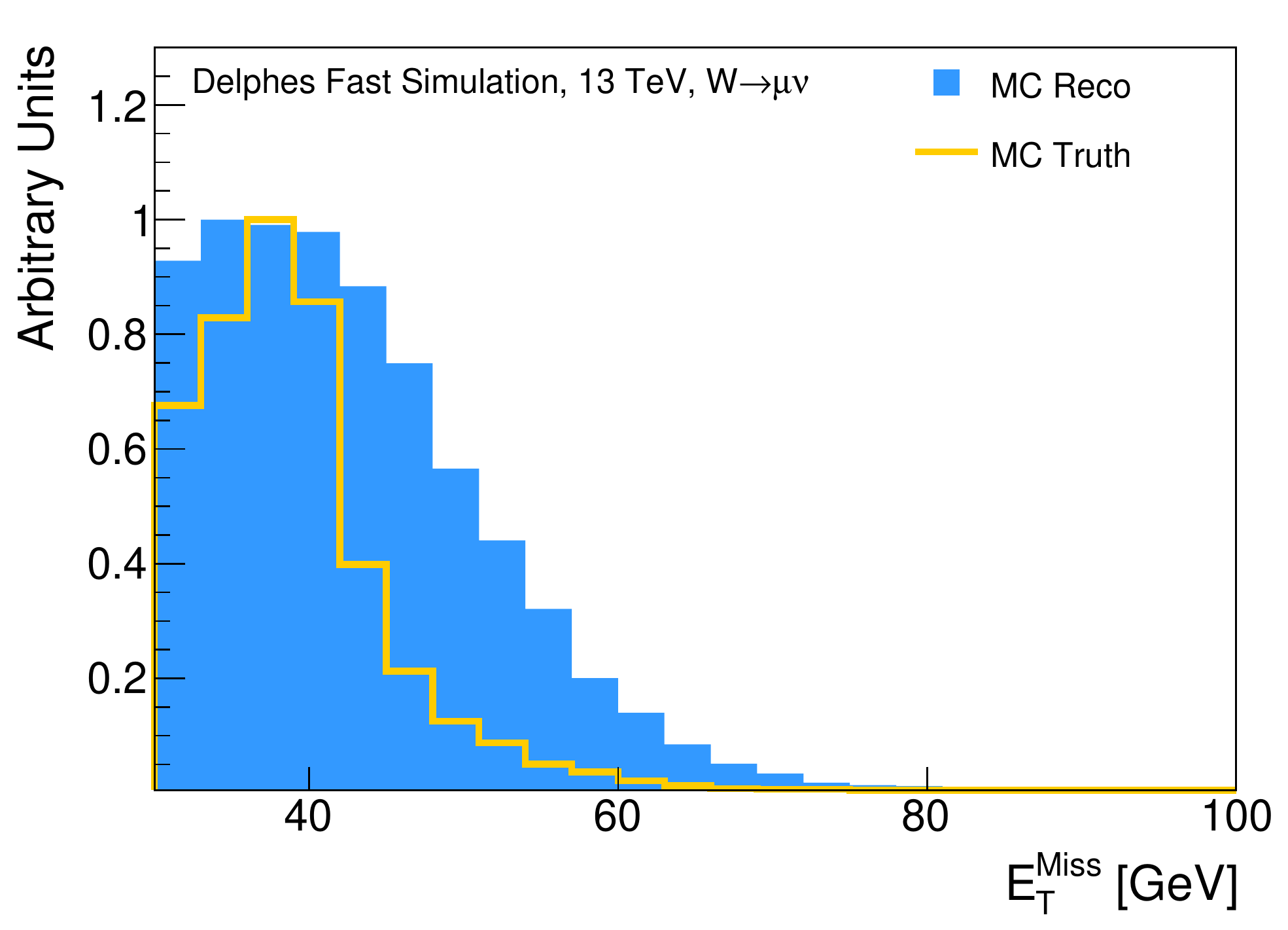} \hspace{0.1cm}
\includegraphics[width=7.3cm]{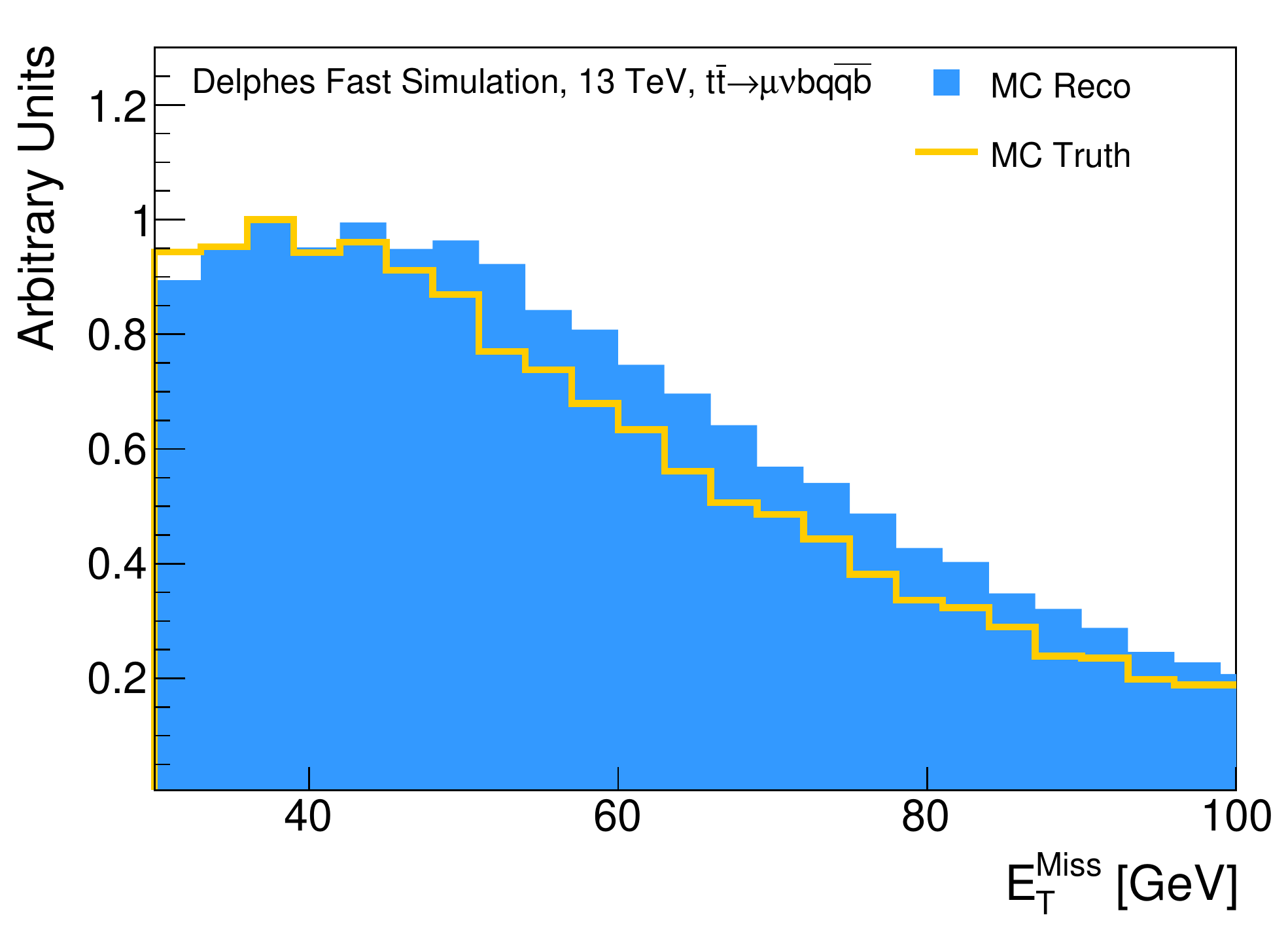}
\caption{\label{fig:ETMissDistributions} Normalized $\MET$  distribution on particle and detector-level for $W\rightarrow\mu\nu$ (left) and $t\bar{t} \rightarrow \mu\nu b q\bar{q}\bar{b}$ (right) events.}
\end{center}
\end{figure}

The situation is different for resolution and migration effects. Figure \ref{fig:ETMissDistributions} shows the reconstructed $\MET$ distribution and the neutrino $\pT$ for leptonic $W$ boson decays as well as semi-leptonic top-quark pair decay within the \textsc{Delphes}-framework. Both distributions indicate significantly larger reconstructed values of  $\MET$  than the underlying particle-level distribution because of the relatively poor $\MET$  resolution and the falling spectra of $p_\text{T}(\nu)>40$\,GeV. A fiducial phase-space definition invoking a minimum $\MET$  value of 60\,GeV will, therefore, lead to more reconstructed events than generated events in the fiducial volume when studying an SM $W$ boson. Differences in the neutrino spectrum between $W$ boson and $t\bar{t}$ processes (Figure \ref{fig:ETMissDistributions}), already produce differences in $C$-factors even for smaller cuts on \MET. Any physics model which has inherently larger values of missing transverse energy, e.g. the decay of a massive $W'$ candidate, will have smaller migration effects from outside the fiducial definition since the majority of events will have $\MET$  values on detector- and particle-level well beyond the 60\,GeV threshold. Hence, the $C$-factor for the $W'$ model is expected to be significantly smaller than that for the SM $W$ boson production. This effect is reduced if the selection is based on leptons, which usually offer a good resolution of the signal kinematic, while if the selection is based on a variable such as MET, its poorer resolution plays a crucial role for many reinterpretations of fiducial cross section measurements.

\begin{figure}[htb]
\begin{center}
\includegraphics[width=7.3cm]{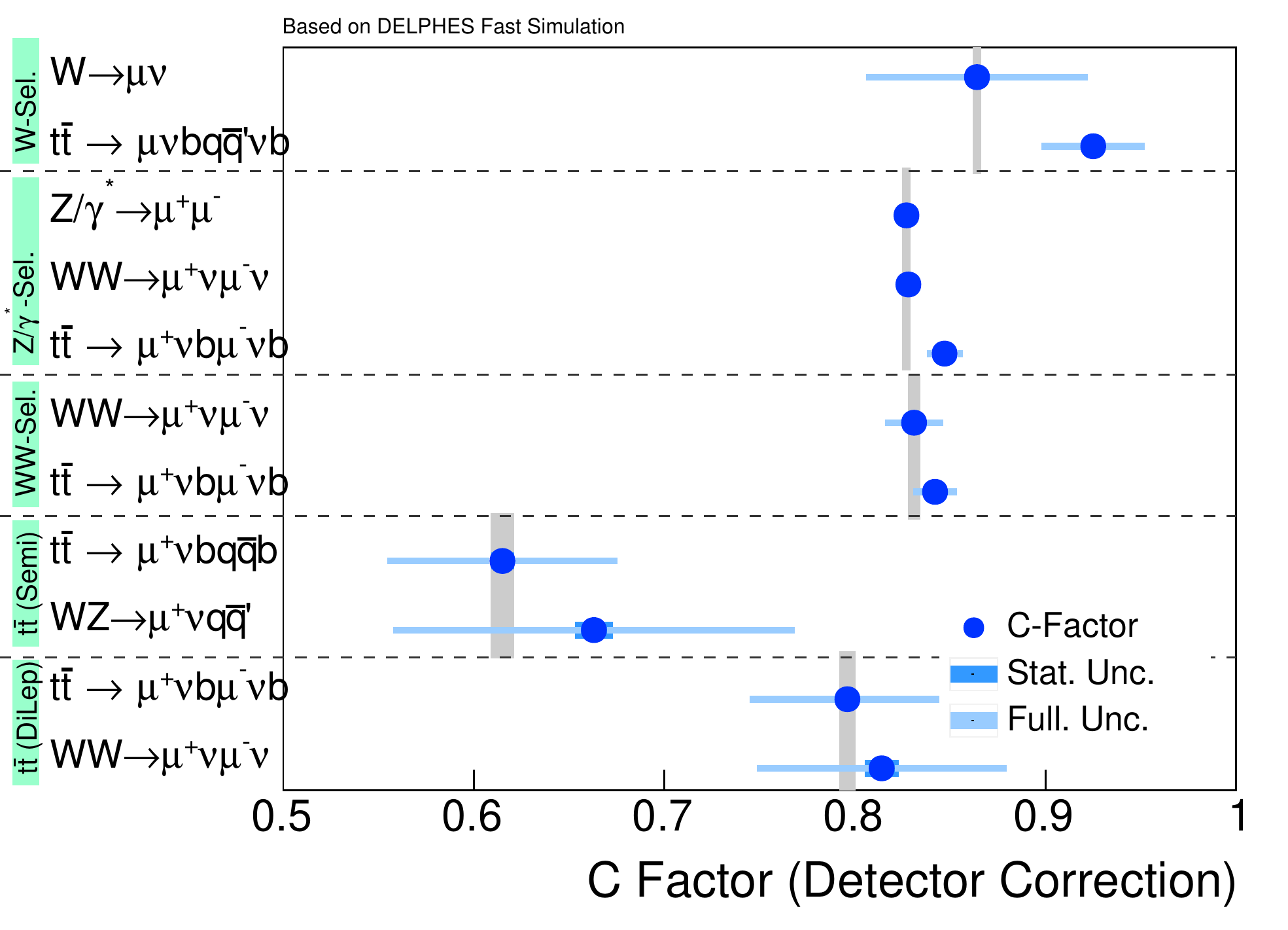} \hspace{0.1cm}
\includegraphics[width=7.3cm]{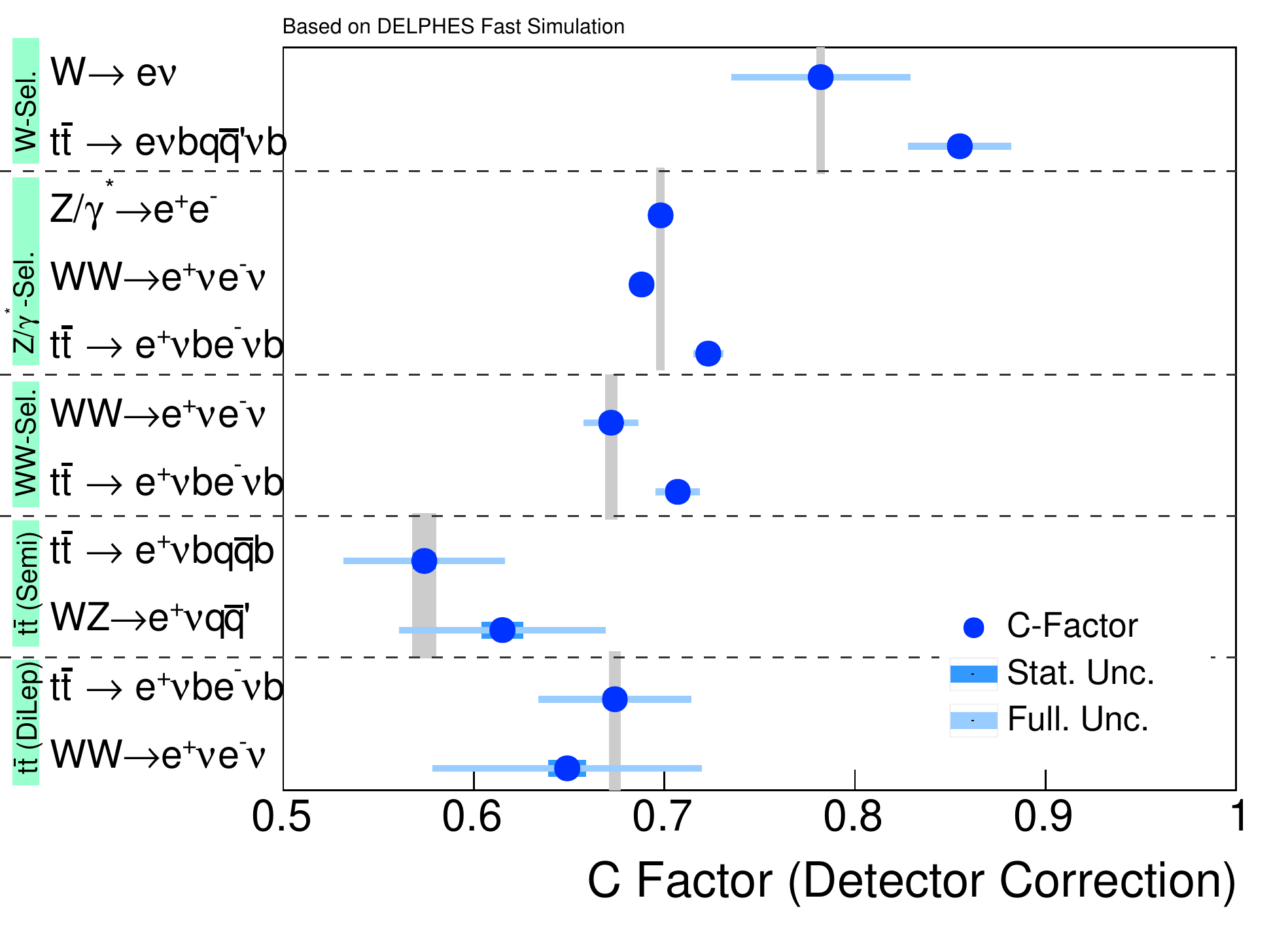}
\caption{\label{fig:CFacSM} Overview of the detector correction $C$-factors for typical SM fiducial regions in the muon decay channel (left) and electron decay channel (right). The first row for each phase-space corresponds to the typical signal process of the chosen phase-space region, the following rows contain $C$-factors of processes that lead to the same final state. The statistical and estimated experimental uncertainty on the $C$-factors is also shown. The gray band indicates the $C$-factor and its uncertainty for the signal process at which the selection is targeted. The systematic uncertainties for processes within one selection are highly correlated.}
\end{center}
\end{figure}

The $C$-factors for SM processes in various fiducial volumes are summarized in Table \ref{tab:SMModelOverview} for electron and muon final states. Selected results are illustrated in Figure \ref{fig:CFacSM}. It should be noted that only the statistical uncertainties should be considered when judging on the differences of these numbers, as the systematic uncertainties are highly correlated. As discussed above, $C$-factors differ when the final state object multiplicities are not equal. Therefore, only processes with the same number of final state objects are compared, e.g. only processes with exactly 2 oppositely charged muons in the final state are compared to each other; i.e. the $Z$ boson decay into two leptons is not compared to the $C$-factors for the $W$ boson selection even though a significant fraction of Z boson events would pass the selection requirements in the fiducial volume, as one lepton might be beyond the detector acceptance. The $C$-factors for all studied SM processes considered in each fiducial volume do not deviate by more than $\approx$10\% from the process for which the fiducial region was designed. The discrepancies result from differences in the $\eta$ distribution of leptons, isolation behaviors of the final state objects, and migration effects of $\MET$ and jet observables. For most processes, the differences noted are on the same level as typical systematic uncertainties on the $C$-factors.

\begin{table}[h]
\center
\footnotesize
\caption{Detector correction $C$-factors for various SM process selections, defined in Table \ref{tab:FidDefinitions}, applied on the corresponding SM signal process in the first rows as well as further SM processes with a similar final state in the following rows. The statistical and estimated experimental uncertainty on the $C$-factors is also indicated. As motivated in Section \ref{sec:ResultSM}, events which did not have the proper number of inclusive particle-level leptons are vetoed.}
{
    \begin{tabular}{l | c|c|c}
\hline
\multicolumn{2}{c}{Muon Decay Channel}											&	\multicolumn{2}{c}{Electron Decay Channel}									\\
\hline
Process										& C$\pm$stat.$\pm$sys.			&	Process									&C$\pm$stat.$\pm$sys.			\\
\hline
\multicolumn{4}{c}{W Selection}	\\
\hline
$W^\pm \rightarrow \mu^\pm \nu$						&	$0.864\pm0.002\pm0.058$	&	$W^\pm \rightarrow e \nu$						&	$0.782\pm0.002\pm0.047$	\\	
$t \bar{t} \rightarrow q \bar{q}' b \bar{b} \mu^\pm \nu $		&	$0.925\pm0.002\pm0.027$	&	$t \bar{t} \rightarrow q \bar{q}' b  \bar{b} e^\pm \nu $	&	$0.855\pm0.002\pm0.027$	\\	
$W^\pm Z\rightarrow \mu^\pm \nu q \bar{q}$					&	$0.938\pm0.001\pm0.034$	&	$W^\pm Z\rightarrow e^\pm \nu q \bar{q}$					&	$0.85\pm0.001\pm0.032$	\\	
$W^+W^-\rightarrow \mu^\pm \nu q \bar{q}'$					&	$0.961\pm0.002\pm0.042$	&	$W^+W^-\rightarrow e^\pm \nu q \bar{q}'$				&	$0.877\pm0.003\pm0.043$	\\	
\hline
\multicolumn{4}{c}{Z Selection}\\
\hline
$Z/\gamma^{*}\rightarrow \mu^{+} \mu^{-}$			&	$0.827\pm0.002\pm0.003$	&	$Z/\gamma^{*}\rightarrow e^{+} e^{-}$			&	$0.698\pm0.002\pm0.003$	\\	
$t \bar{t} \rightarrow \mu^+ \nu b + \mu^- \nu \bar{b}$		&	$0.847\pm0.005\pm0.008$	&	$t \bar{t} \rightarrow e^+ \nu b + e^- \nu \bar{b}$	&	$0.723\pm0.005\pm0.006$	\\	
$W^{+}W^{-}\rightarrow \mu^{+}\nu \mu^{-} \nu$				&	$0.828\pm0.003\pm0.003$	&	$W^{+}W^{-}\rightarrow e^{+}\nu e^{-} \nu$			&	$0.688\pm0.003\pm0.003$	\\	\hline
\multicolumn{4}{c}{WW Selection}\\
\hline
$W^{+}W^{-}\rightarrow l^{+}\nu l^{-} \nu$				&	$0.831\pm0.003\pm0.015$	&	$W^{+}W^{-}\rightarrow l^{+}\nu l^{-} \nu$			&	$0.672\pm0.003\pm0.014$		\\
$t \bar{t} \rightarrow \mu^+ \nu b + \mu^- \nu \bar{b}$		&	$0.842\pm0.003\pm0.011$	&	$t \bar{t} \rightarrow e^+ \nu b + e^- \nu \bar{b}$	&	$0.707\pm0.004\pm0.011$	\\	
\hline
\multicolumn{4}{c}{Top-Pair Selection (di-lep.)}\\
\hline
$t \bar{t} \rightarrow \mu^+ \nu b + \mu^- \nu \bar{b}$		&	$0.796\pm0.004\pm0.051$	&	$t \bar{t} \rightarrow e^+ \nu b + e^- \nu \bar{b}$	&	$0.674\pm0.003\pm0.040$	\\	
$W^{+}W^{-}\rightarrow l^{+}\nu l^{-} \nu$				&	$0.814\pm0.009\pm0.065$	&	$W^{+}W^{-}\rightarrow l^{+}\nu l^{-} \nu$			&	$0.649\pm0.010\pm0.070$	\\	
\hline
\multicolumn{4}{c}{Top-Pair Selection (semi-lep.)}\\
\hline
$t \bar{t} \rightarrow q \bar{q}' b \bar{b} \mu^\pm \nu $		&	$0.615\pm0.006\pm0.060$	&	$t \bar{t} \rightarrow q \bar{q}' b \bar{b} e^\pm \nu $	&	$0.574\pm0.006\pm0.042$		\\
$W^\pm Z\rightarrow \mu^{\pm} \nu f \bar{f}$				&	$0.663\pm0.01\pm0.105$	&	$W^{\pm}Z\rightarrow e^{\pm} \nu f \bar{f}$			&	$0.615\pm0.011\pm0.053$	\\	
\hline
\multicolumn{4}{c}{WZ Selection}\\
\hline
$W\pm Z\rightarrow \mu^\pm \nu \mu^{+}\mu^{-} (LO)$				&	$0.736\pm0.003\pm0.006$	&	$W^\pm Z\rightarrow e^\pm \nu e^{+}e^{-} (LO)$			&	$0.559\pm0.003\pm0.003$		\\
$W\pm Z\rightarrow \mu\pm \nu \mu^{+}\mu^{-}$				&	$0.740\pm0.002\pm0.004$	&	$W^\pm Z\rightarrow e^\pm \nu e^{+}e^{-}$			&	$0.560\pm0.002\pm0.003$	\\
\hline
\end{tabular}

\label{tab:SMModelOverview}
}
\end{table}

\subsection{Reinterpretation with Effective Field Theories\label{sec:ResultEFT}}

\begin{figure}[htb]
\begin{center}
\includegraphics[width=7.3cm]{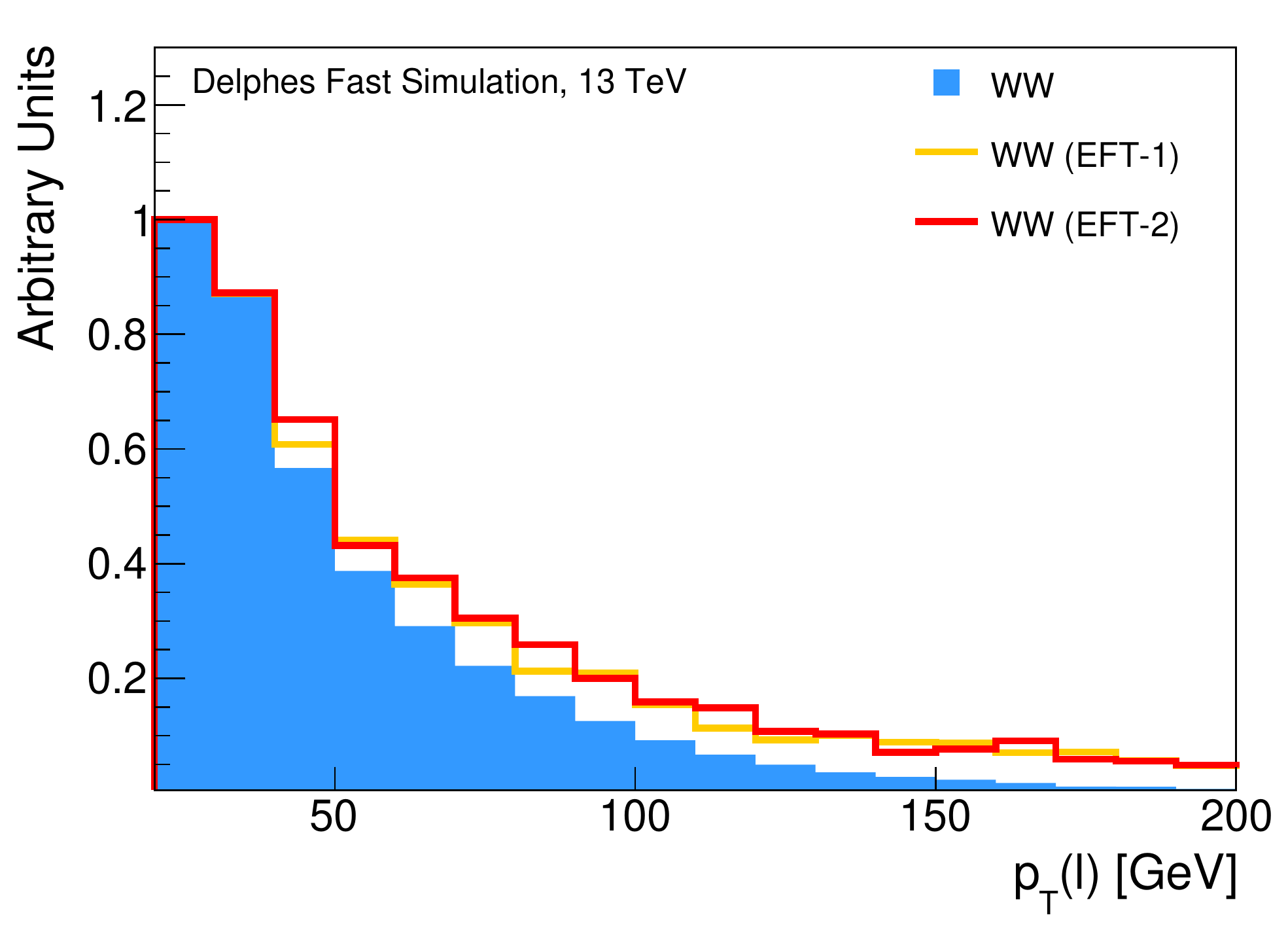} \hspace{0.1cm}
\includegraphics[width=7.3cm]{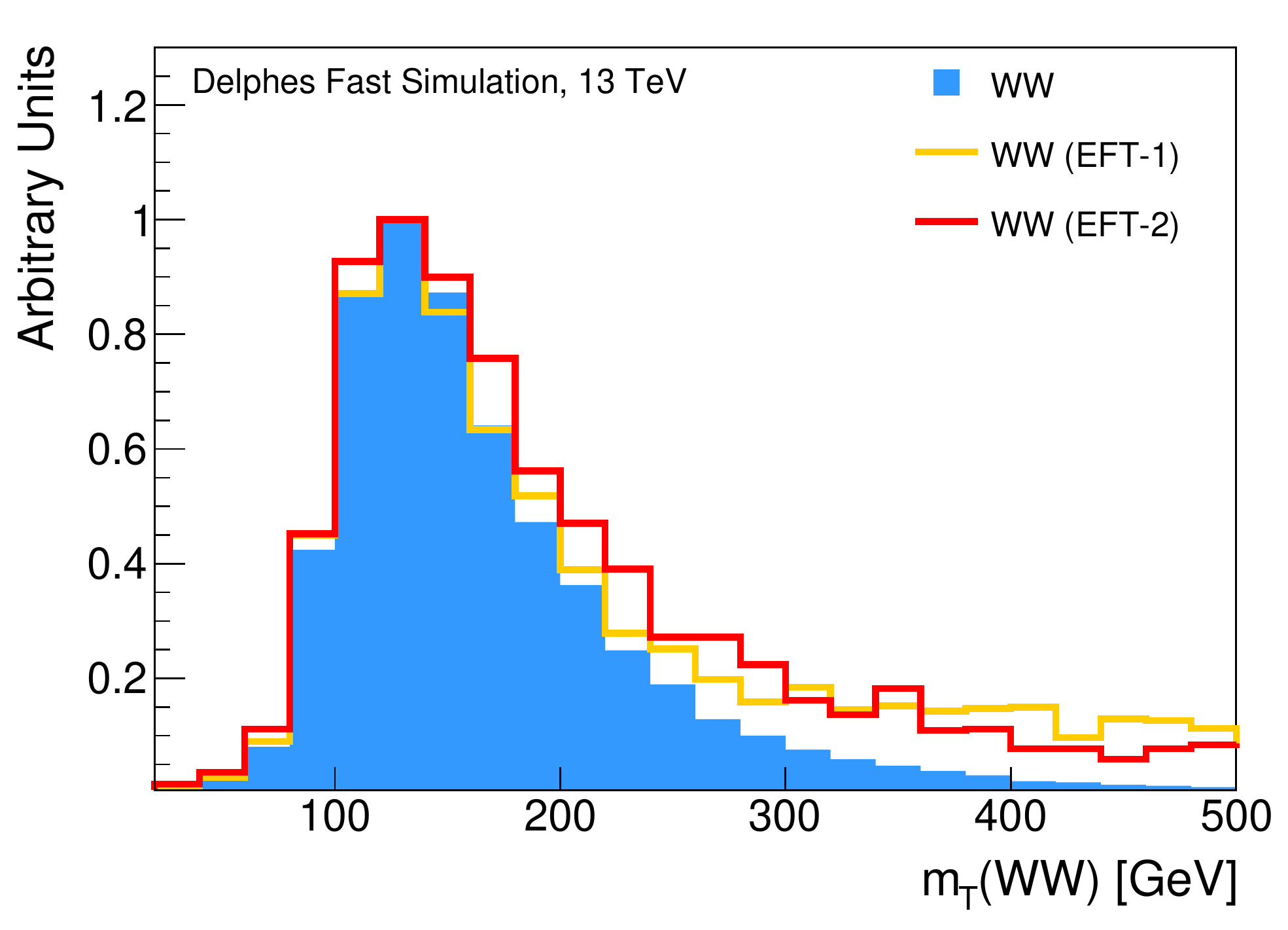}
    \caption{\label{fig:EFTIl} Leading lepton $\pT$ spectrum (left) and diboson transverse mass distribution, $m_\text{T}(WW)$, (right) for the $pp\rightarrow WW\rightarrow l\nu\nu$ process with the SM prediction (at LO in $\alpha_s$) as well as two EFT parameter choices.}
\end{center}
\end{figure}

While most direct searches aim for the observation of new resonances, dim-6 operators of EFTs impact the high energy tails of SM process distributions, such as the invariant mass of diboson final states or the transverse momentum of decay leptons. It is important to note that the effect of these operators mainly changes the kinematics of the SM process, and thus the kinematics of the decay products, while the number of final state objects remains constant. Since the effects of EFT operators exhibit a large energy dependence, they are typically studied using differential cross sections as a function of an energy-dependent observable. In the following, we investigate the impact of two BSM EFT operator choices on the $C$-factors in a sensitive fiducial volume. The first parameter choice (EFT-1) is $c_{WWW}/\Lambda^2$ = -35 implemented in \textsc{MadGraph EWdim6}, the second (EFT-2) $c_{W}/\Lambda^2$ = 40 in the same model. Figure \ref{fig:EFTIl} illustrates the impact of these model parameters on WW production in proton-proton collisions in the leptonic decay channel. The leading lepton $\pT$ spectrum, as well as the diboson transverse mass distribution $m_\text{T}(WW)$ (Eq. \ref{eqn:mt} for $i=2$) are both enhanced at large values compared to the SM prediction. Hence, typical limits on EFT operators are derived in fiducial phase-space regions which test the high energy tails of differential distributions. We study two fiducial volumes in the $WW$ and $WZ$ boson production by modifying the standard SM selection for $WW$ and $WZ$ processes. First, a minimal cut on the $\pT$ of the leading lepton of 100 and 80\,GeV is tested, then a minimal cut on the diboson transverse mass of  200 and 250\,GeV (Table \ref{tab:FidDefinitions}) is examined.

\begin{figure}[htb]
\begin{center}
\includegraphics[width=7.3cm]{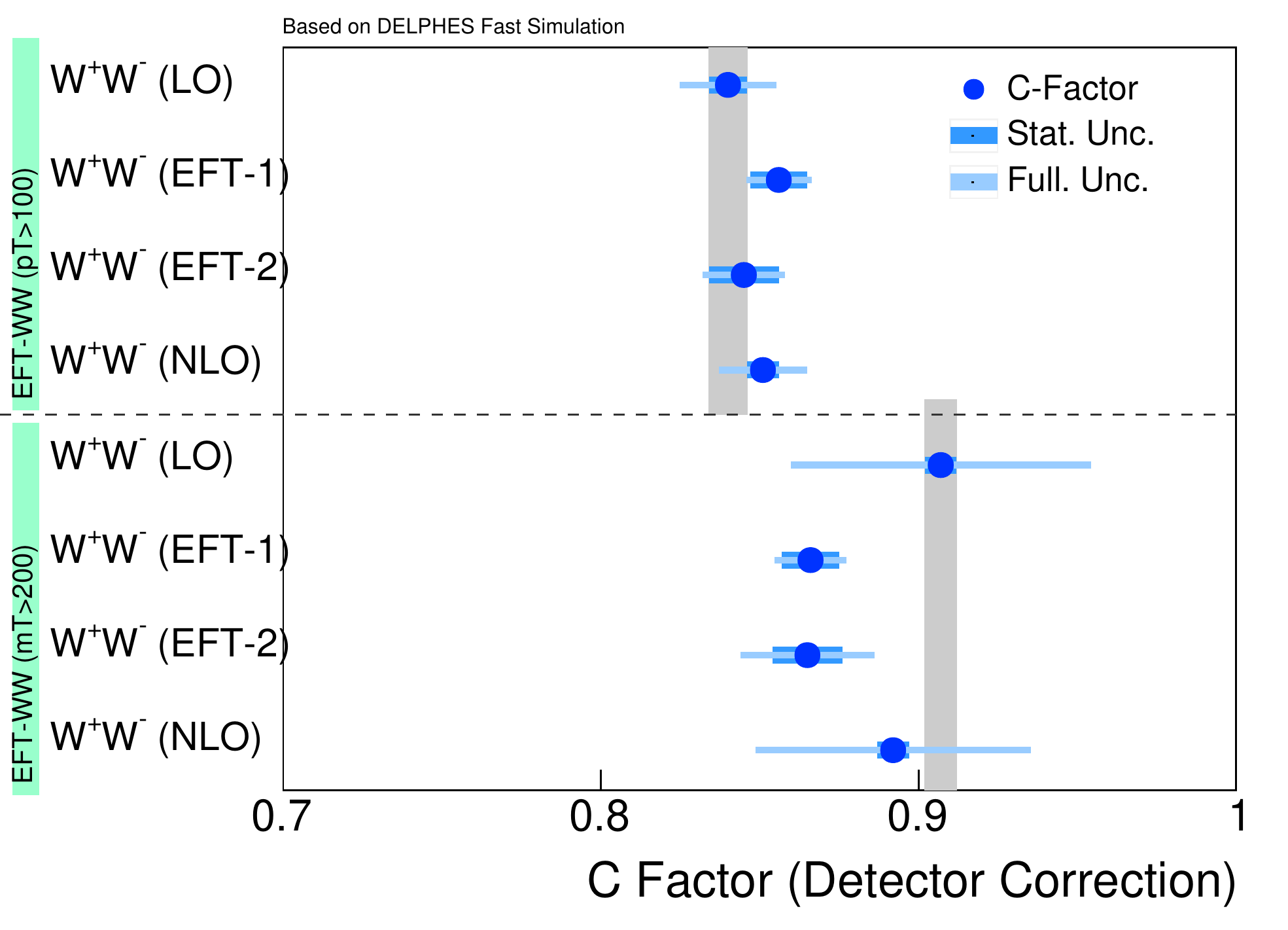} \hspace{0.1cm}
\includegraphics[width=7.3cm]{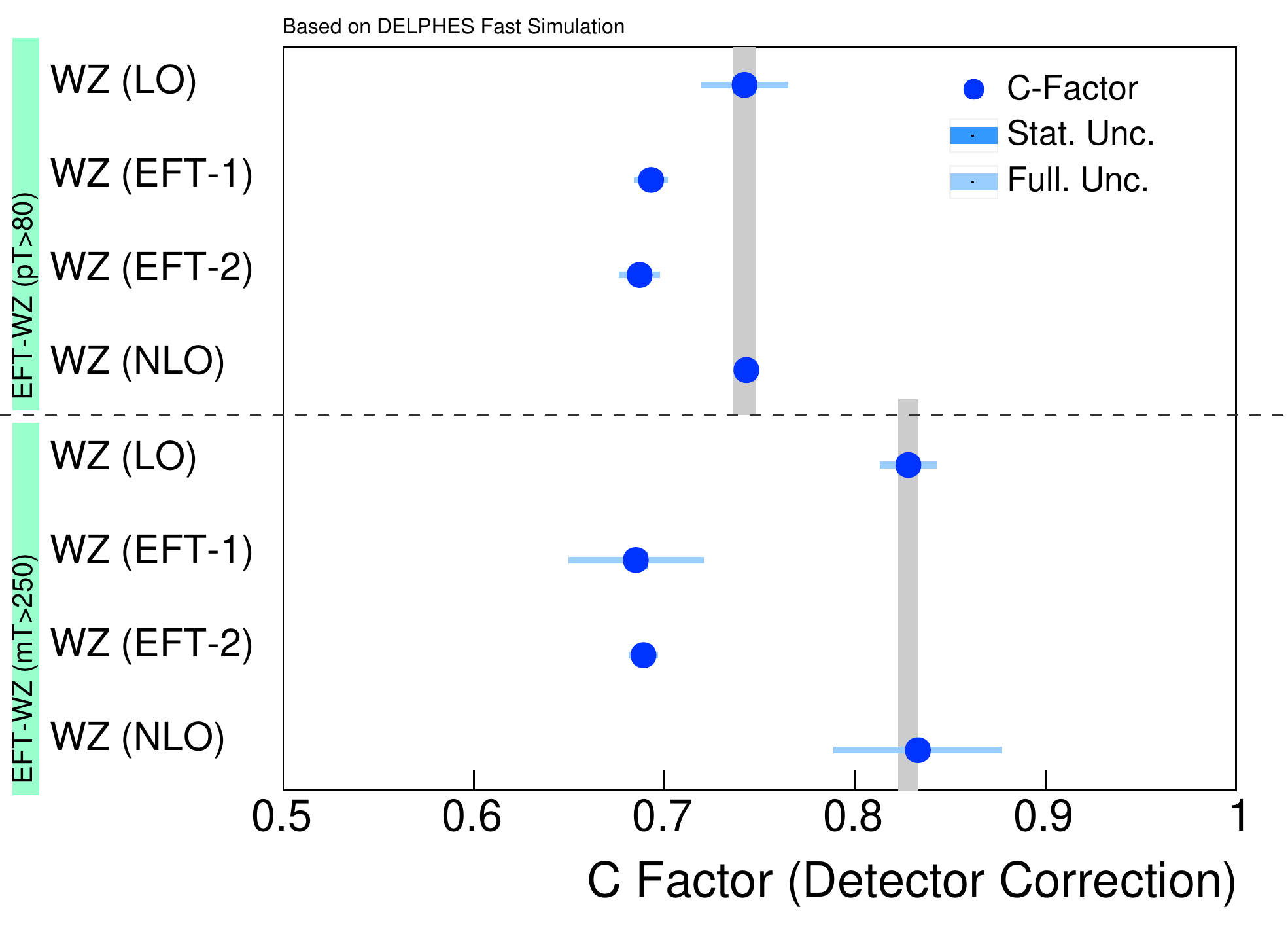}
\caption{\label{fig:CFacEFTSM} Overview of detector correction $C$ factors for two different EFT parameter choices for $WW$ (left) and $WZ$ (right) production with selection cuts on the leading lepton $\pT$ (upper half) and the diboson transverse mass (lower half), defined in Table \ref{tab:FidDefinitions}. The statistical and estimated experimental uncertainty on the C-factors is also indicated. The gray band indicates the C-factor and its uncertainty for the signal process at which the selection is targeted. The systematic uncertainties for processes within one selection are highly correlated.}
\end{center}
\end{figure}

The resulting $C$-factors for the $WW$ and $WZ$ diboson production for both EFT scenarios are summarized in Table \ref{tab:EFTModelOverview} and illustrated in Figure \ref{fig:CFacEFTSM}. As stated in Section~\ref{Sec:Samples}, the leading-order predictions in $\alpha_s$ have been used for both the SM and EFT prediction. The expected $C$-factors for the NLO SM prediction are also shown for comparison. While the $C$-factors for EFT-sensitive fiducial volumes for the $WW$ diboson production, defined by a cut on the $\pT$ of the leading lepton, show a good agreement between the SM prediction and the tested EFT models, we observe deviations up to 5\% for phase-space regions that are defined by a requirement on \mT. The cut-value on \mT  is so large that relevant migration effects are caused by the degradation of MET resolution, due to the significantly higher adronic activity present in the tested EFT models. When assuming a perfect reconstruction of $\MET$, the differences vanish. We observe significant differences for both EFT-sensitive fiducial volumes in the $WZ$ final state. These are again caused by the significantly higher hadronic activity caused by our EFT-parameter choice which leads to a reduction of events that pass the isolation criteria on the leptons. Since three isolated leptons are required, this effect is amplified compared to single- or di-lepton final states.

\begin{table}[h]
\footnotesize
\center
\caption{Detector correction $C$ factors for two different EFT parameter choices for $WW$ (left) and $WZ$ (right) production with selection cuts on the leading lepton $\pT$ (EFT-Sensitive Selection 1) and the diboson transverse mass (EFT-Sensitive Selection 2), defined in Table \ref{tab:FidDefinitions}. The statistical and estimated experimental uncertainty on the $C$-factors is also indicated. As motivated in section \ref{sec:ResultSM}, events which did not have the proper number of inclusive particle-level leptons are vetoed.}
{
    \begin{tabular}{l | c|c|c}
\hline
WW-Final State								& C$\pm$stat.$\pm$sys.			&	WZ-Final State								&C$\pm$stat.$\pm$sys.			\\
\hline
\multicolumn{2}{c|}{EFT-Sensitive Selection 1:}									& 	\multicolumn{2}{c}{EFT-Sensitive Selection 1: } 	\\
\hline
$W^{+}W^{-}\rightarrow \mu^{+}\nu \mu^{-} \nu$\,(LO)		&	$0.84\pm0.006\pm0.014$	&	$W^\pm Z\rightarrow \mu^\pm \nu \mu^{+}\mu^{-}$\,(LO)	&	$0.742\pm0.006\pm0.022$ \\
$W^{+}W^{-}\rightarrow \mu^{+}\nu \mu^{-} \nu$\,(EFT-1)		&	$0.856\pm0.009\pm0.005$	&	$W^\pm Z\rightarrow \mu^\pm \nu \mu^{+}\mu^{-}$\,(EFT-1)	&	$0.693\pm0.004\pm0.008$	\\
$W^{+}W^{-}\rightarrow \mu^{+}\nu \mu^{-} \nu$\,(EFT-2)		&	$0.845\pm0.011\pm0.005$	&	$W^\pm Z\rightarrow \mu^\pm \nu \mu^{+}\mu^{-}$\,(EFT-2)	&	$0.687\pm0.004\pm0.01$	\\	
$W^{+}W^{-}\rightarrow \mu^{+}\nu \mu^{-} \nu$\,(NLO)		&	$0.851\pm0.005\pm0.013$	&	$W^\pm Z\rightarrow \mu^\pm \nu \mu^{+}\mu^{-}$\,(NLO)		&	$0.743\pm0.003\pm0.005$	\\	
\hline
\multicolumn{2}{c|}{EFT-Sensitive Selection 2:}									& 	\multicolumn{2}{c}{EFT-Sensitive Selection 2: } 	\\
\hline
$W^{+}W^{-}\rightarrow \mu^{+}\nu \mu^{-} \nu\,$(LO)		&	$0.907\pm0.005\pm0.047$	&	$W^\pm Z\rightarrow \mu^\pm \nu \mu^{+}\mu^{-}$\,(LO)	&	$0.828\pm0.005\pm0.014$	\\	
$W^{+}W^{-}\rightarrow \mu^{+}\nu \mu^{-} \nu\,$(EFT-1)		&	$0.866\pm0.009\pm0.007$	&	$W^\pm Z\rightarrow \mu^\pm \nu \mu^{+}\mu^{-}$\,(EFT-1)&	$0.685\pm0.006\pm0.035$	\\	
$W^{+}W^{-}\rightarrow \mu^{+}\nu \mu^{-} \nu\,$(EFT-2)		&	$0.865\pm0.011\pm0.018$	&	$W^\pm Z\rightarrow \mu^\pm \nu \mu^{+}\mu^{-}$\,(EFT-2)&	$0.689\pm0.005\pm0.006$	\\	
$W^{+}W^{-}\rightarrow \mu^{+}\nu \mu^{-} \nu\,$(NLO)		&	$0.892\pm0.005\pm0.043$	&	$W^\pm Z\rightarrow \mu^\pm \nu \mu^{+}\mu^{-}$\,(NLO)	&	$0.833\pm0.005\pm0.044$	\\	
\hline
\end{tabular}

\label{tab:EFTModelOverview}
}
\end{table}

\subsection{BSM Search Regions\label{sec:ResultBSM}}

The selections applied for the various BSM models, summarized in Table \ref{tab:FidDefinitions}, probe different potential sources of model dependencies. While the selection aiming at $Z'$ models only involves cuts on leptons, the search for $W'$ adds also a requirement on $\MET$. The search for Leptoquarks models combines selection criteria on leptons as well as jets, while the selections aiming at SUSY and $4^{th}$-generation models targets all major observables, i.e. leptons, $\MET$ as well as (b-)jets.

\begin{figure}[htb]
\begin{center}
\includegraphics[width=7.3cm]{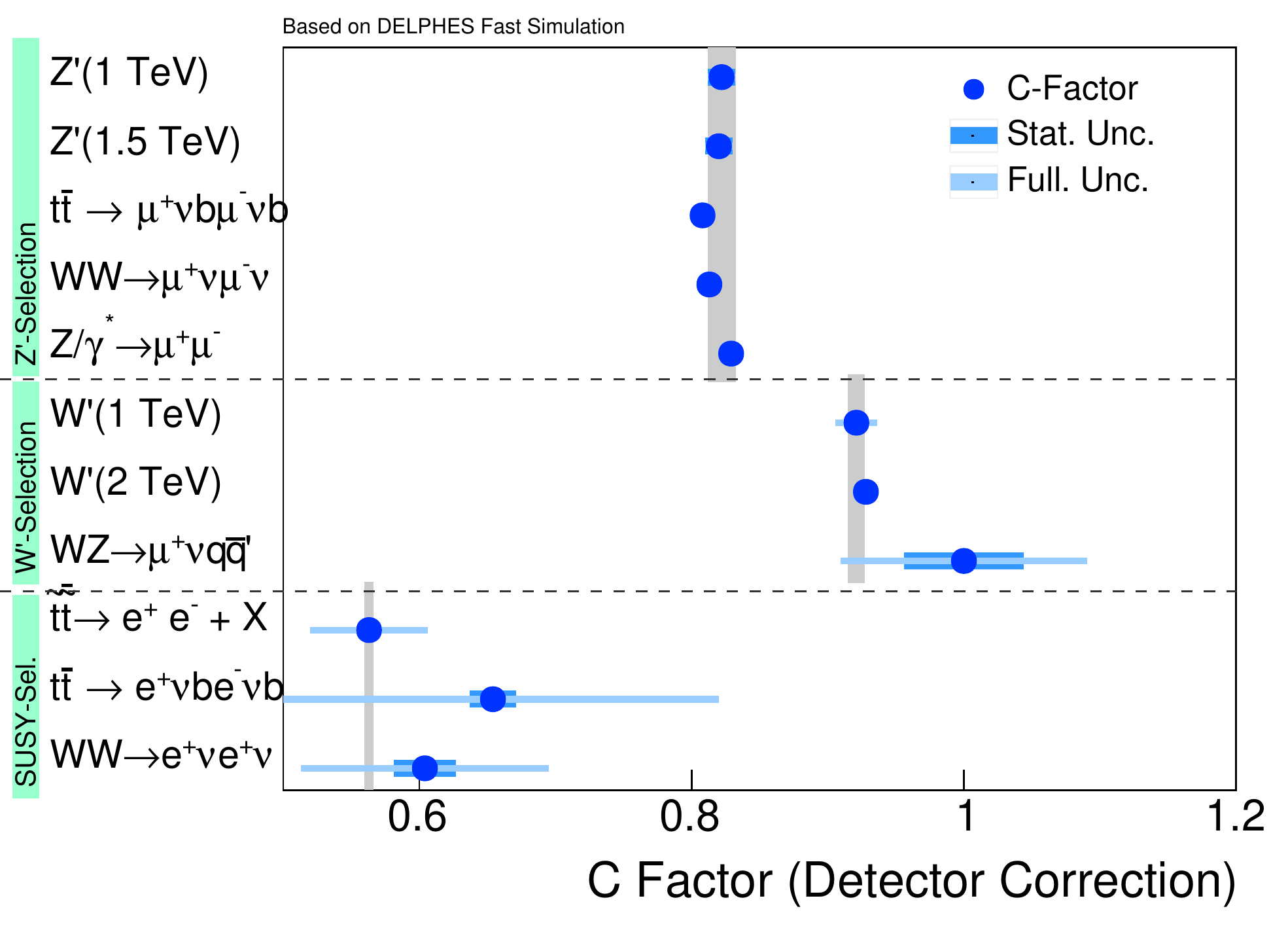} \hspace{0.1cm}
\includegraphics[width=7.3cm]{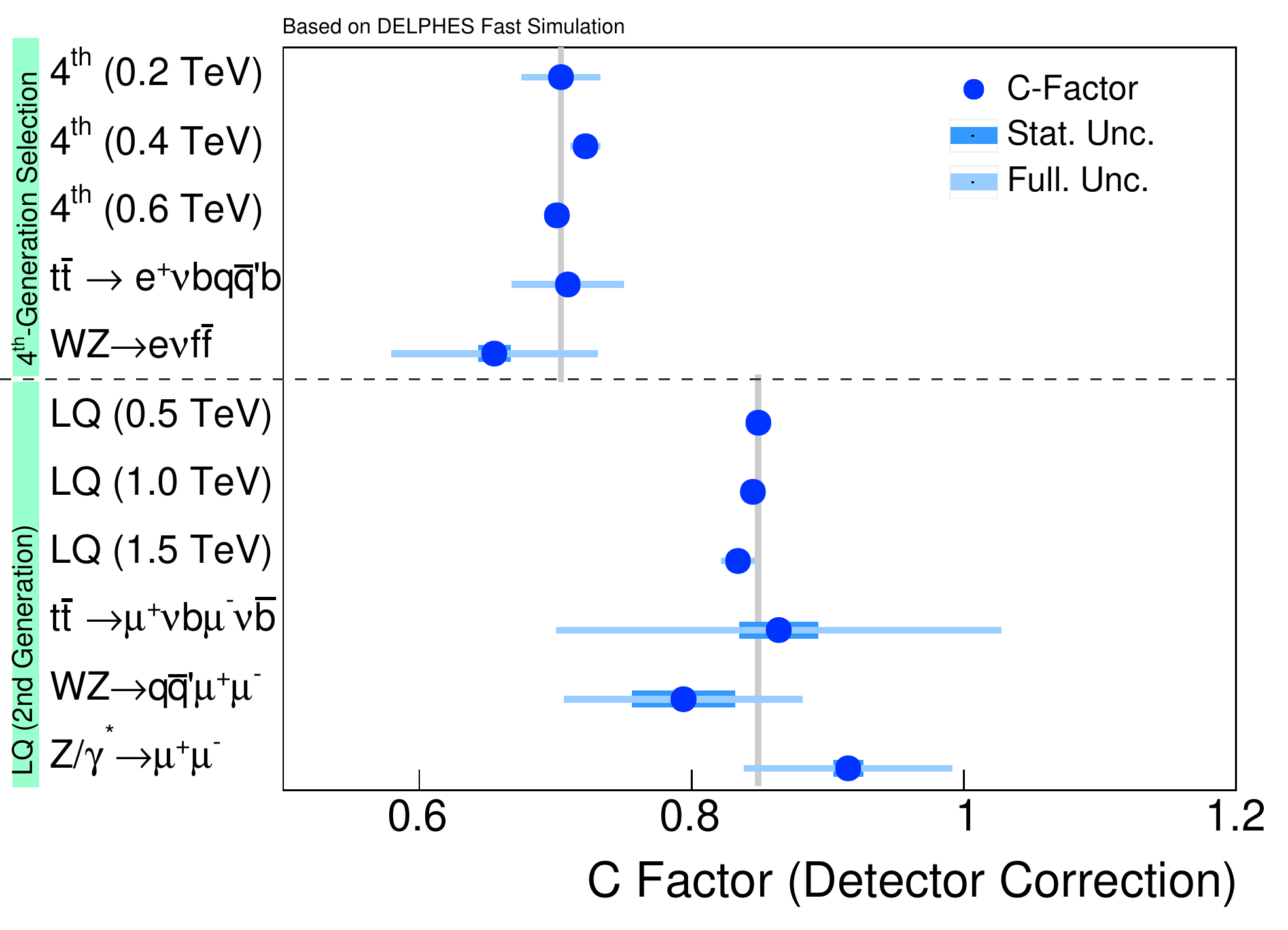}
\caption{\label{fig:CFacBSM} Overview of the detector correction $C$ factors for various BSM Model selections, defined in Table \ref{tab:FidDefinitions}, for the BSM signal processes in the first rows as well as SM processes with a similar final state in the following rows. The statistical and estimated experimental uncertainty on the C-factors is also indicated. The gray band indicates the C-factor and its uncertainty for the signal process at which the selection is targeted.}
\end{center}
\end{figure}

The observed variations of $C$-factors for various BSM processes in the extreme phase-space regions used in direct searches are summarized in Figure \ref{fig:CFacBSM} and Table \ref{tab:ExoticModelOverview}. The $C$-factors of several SM processes with the same final state and similar energies to the BSM signature have also been studied. Model parameters for each BSM model have been varied individually, while the phase-space region remained unchanged. The $C$-factor dependence on BSM model parameters, for a given phase-space region, is minimal as long as the cuts that define the search region are place far enough from the new particle masses, i.e. no threshold effects are expected.  For example, a region defined to search for a $Z'$ model might employ a $m_{ll}$ cut of 500\,GeV instead of the 200\,GeV cut, which is used in our study. The $C$-factors for all models with $m_{Z'}>600$\,GeV will be similar as most BSM events would be far from the phase-space edge. However, for a model with $m_{Z'}=450$\,GeV or $m_{Z'}=550$\,GeV, the corresponding $C$-factors will be much smaller compared due to threshold cut on $m_{ll}$. 

\begin{figure}[htb]
\begin{center}
\includegraphics[width=7.3cm]{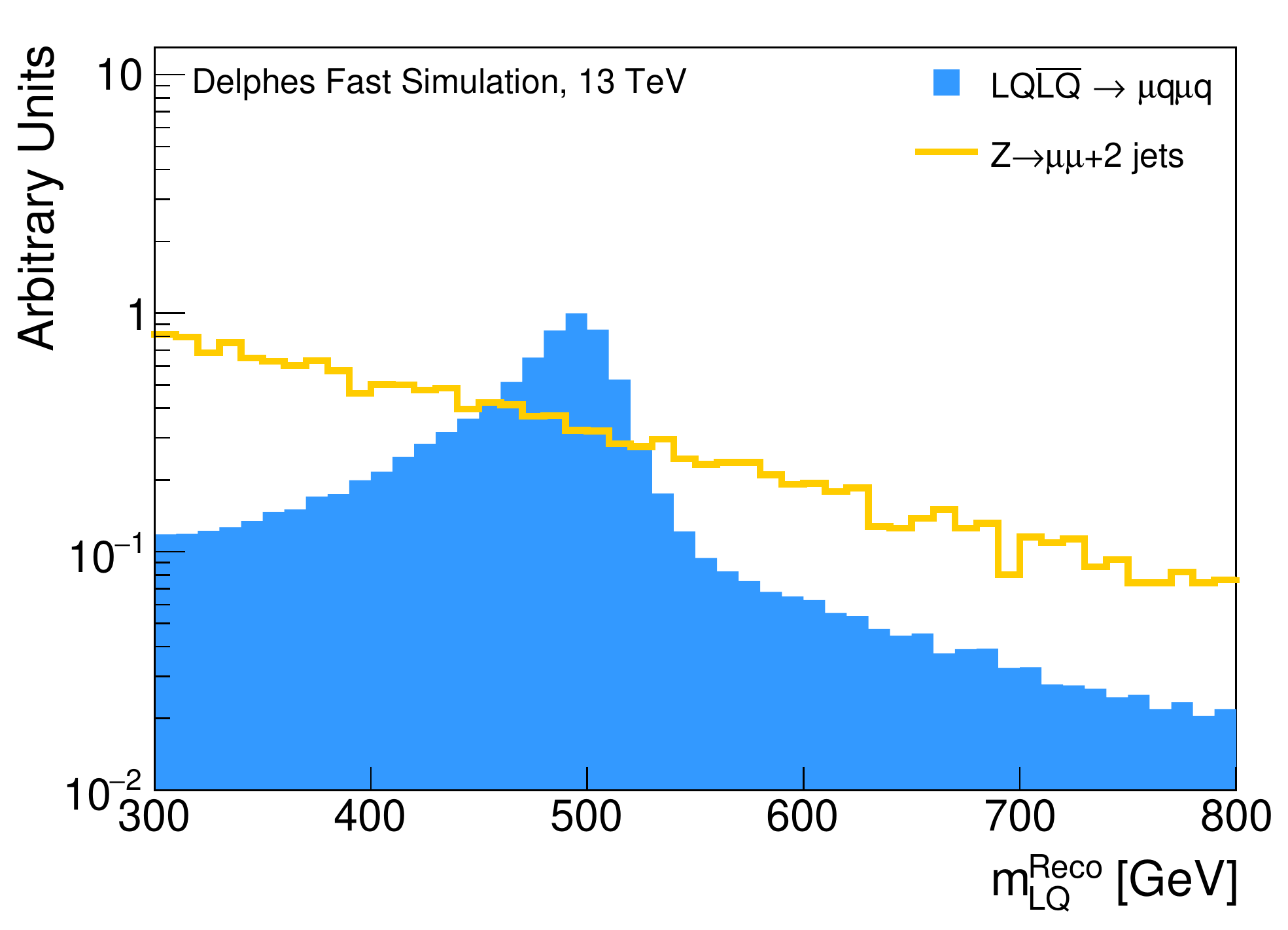} \hspace{0.1cm}
\includegraphics[width=7.3cm]{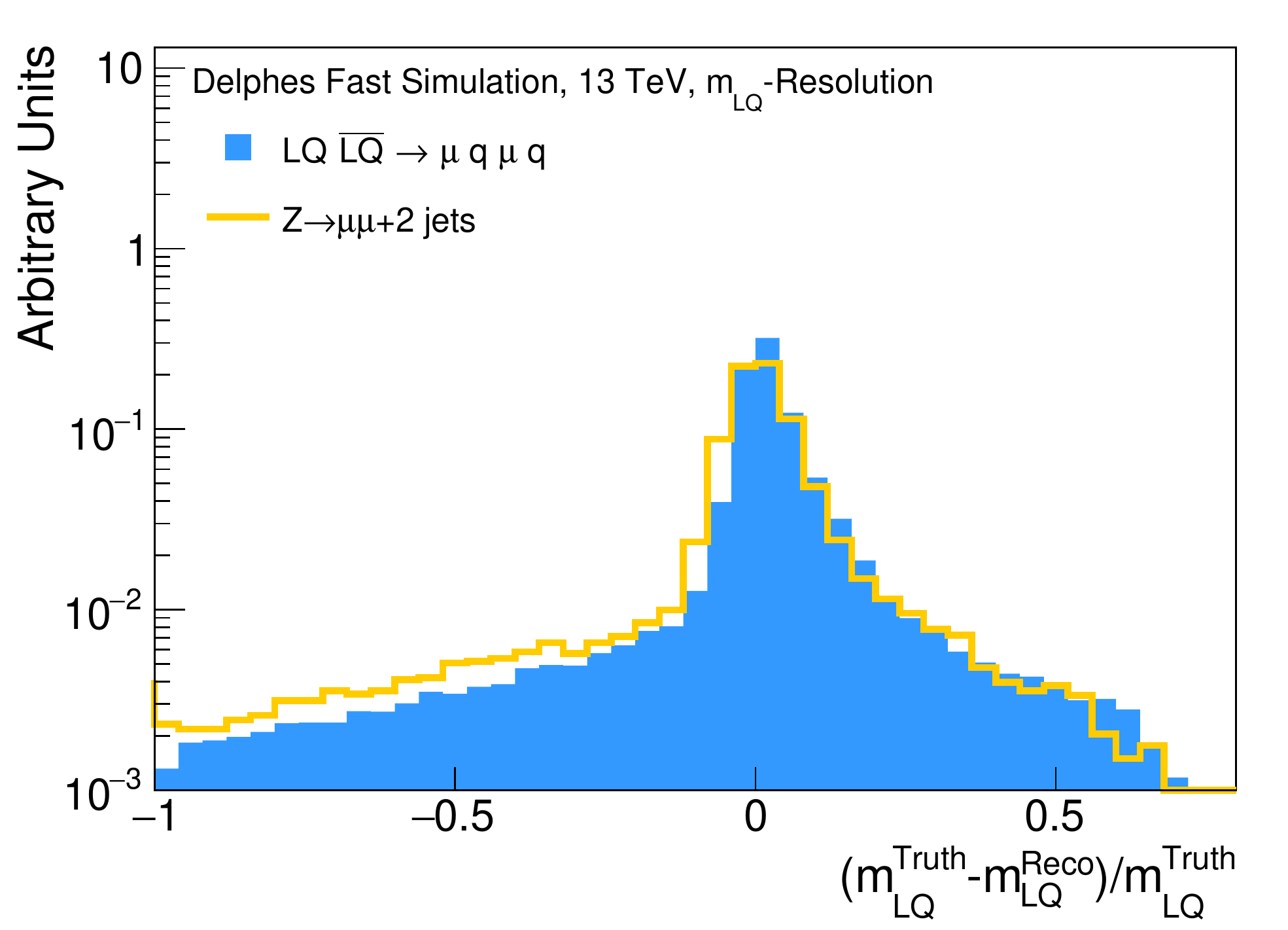}
\caption{\label{fig:DisLQ} Reconstructed mass distribution of selected leptoquark candidates for a signal process with $m_{LQ}=400$\,GeV and the $Z\rightarrow \mu\mu$ process (left) and the mass resolution for both processes (right).}
\end{center}
\end{figure}

All processes passing the $Z'$ selection, defined only by requirements on lepton kinematics, lead to similar $C$-factors due to small migration effects. The situation is different for the selection of potential $W'$ candidates and selections that SUSY, as $\MET$ migration effects become important. In these cases, the reconstructed $\MET$ has larger tails than the particle level $\MET$ observable, leading to more reconstructed events to pass the selection. The search for $4^{th}$-generation models employs cuts on lepton kinematics, $\MET$ and jets, hence we observe convoluted migration effects due to the differences in the $\MET$ and jet kinematics on reconstruction and truth level.
A special case is the search for LQ models which does not involve any $\MET$ related observables and rather consistent $C$-factors are observed. A notable exception is the $C$-factor for the Drell-Yan processes, were a statistical significant difference can be seen. This difference can be traced back to the requirement on the lepto-quark candidate mass, $m_{LQ}$, which is defined as the invariant mass between one lepton and one jet, shown for a LQ signal sample and the $Z\rightarrow\mu\mu$ processes in Figure \ref{fig:DisLQ}. While the distributions are clearly very different, naively no significant effect on the $C$-factor is expected, as the cut on $m_{LQ}$ is applied particle- and detector level. However, when looking at the  resolution of the $m_{LQ}$ observable, significantly larger trails towards higher reconstructed masses become visible (Figure \ref{fig:DisLQ}). These one-sided tails lead therefore to similar migrations effects as have been observed for \MET. 

In summary, the studied selections lead to deviations of the C factors by up to 20\%. These deviations are mainly caused by $\MET$ requirements in the definition of the fiducial space-phase; however, potentially all observables with asymmetric tails can lead to significant migration effects. In fact, it was already shown in reference\cite{Aad:2019fac}, targeting the search for Z’ and W’, that not only such resolution effects, but also the lepton identification itself could lead to very significant model-dependent selection efficiencies.

 \begin{table}[h]
 \footnotesize
\center
\caption{Detector correction $C$ factors for various BSM Model selections, defined in Table \ref{tab:FidDefinitions}, for the BSM signal processes in the first rows as well as SM processes with a similar final state in the following rows. The statistical and estimated experimental uncertainty on the $C$-factors is also indicated. As motivated in section \ref{sec:ResultSM}, events which did not have the proper number of inclusive particle-level leptons are vetoed.}
{
\begin{tabular}{l | c|c|c}
\hline
Process											& C$\pm$stat.$\pm$sys.			&	Process											&C$\pm$stat.$\pm$sys.\\
\hline
\multicolumn{2}{c|}{\underline{$ Z'$ Selection ($\mu$):$$}}							& 	\multicolumn{2}{c}{\underline{$ W'$ Selection ($\mu$): $$} } 	\\
$Z'(500\,GeV)$										&	$0.822\pm0.011\pm0.001$	&	$W'(1000\,GeV)$									&	$0.921\pm0.006\pm0.014$	\\
$Z'(1000\,GeV)$									&	$0.822\pm0.01\pm0.001$	&	$W'(1500\,GeV)$									&	$0.928\pm0.006\pm0.007$ \\	
$Z'(1500\,GeV)$									&	$0.82\pm0.01\pm0.001$	&	$W'(2000\,GeV)$									&	$0.931\pm0.006\pm0.006$	\\	
$t \bar{t} \rightarrow \mu^+ \nu b + \mu^- \nu \bar{b}$			&	$0.808\pm0.007\pm0.005$	&	$W^{+}Z\rightarrow l^{+} \nu f \bar{f}$					&	$1.000\pm0.044\pm0.079$	\\	
$W^{+}W^{-}\rightarrow l^{+}\nu l^{-} \nu$					&	$0.813\pm0.004\pm0.001$	&	\multicolumn{2}{c}{\underline{ SUSY Selection (e): }} 	\\
$Z/\gamma^{*}\rightarrow \mu^{+} \mu^{-}\,^{(m_{ll}>200\,GeV)}$	&	$0.803\pm0.002\pm0.001$	&	$\tilde{t}\bar{\tilde{t}}\rightarrow e^+ e^- + X$				&	$0.563\pm0.003\pm0.043$	\\	
$Z/\gamma^{*}\rightarrow \mu^{+} \mu^{-}\,^{(m_{ll}>500\,GeV)}$	&	$0.829\pm0.002\pm0.001$	&	$t \bar{t} \rightarrow e^+ \nu b + e^- \nu \bar{b}$			&	$0.654\pm0.017\pm0.165$	\\	
\multicolumn{2}{c|}{\underline{$$ LQ Selection ($\mu$): $$ }} 			&	$W^{+}W^{-}\rightarrow l^{+}\nu l^{-} \nu$					&	$0.604\pm0.023\pm0.088$	\\	
$LQ(2nd\,Gen,\,1000\,GeV)$							&	$0.849\pm0.002\pm0.007$	&	\multicolumn{2}{c}{\underline{ 4th Gen. Selection (e):  }} 	\\
$LQ(2nd\,Gen,\,1500\,GeV)$							&	$0.845\pm0.002\pm0.007$	&	$4^{th}\,Gen. (200\,GeV)$								&	$0.704\pm0.002\pm0.029$	\\	
$LQ(2nd\,Gen,\,500\,GeV)$							&	$0.834\pm0.003\pm0.012$	&	$4^{th}\,Gen. (400\, GeV)$								&	$0.722\pm0.002\pm0.011$	\\	
$t \bar{t} \rightarrow \mu^+ \nu b + \mu^- \nu \bar{b}$			&	$0.864\pm0.029\pm0.161$	&	$4^{th}\,Gen. (600\,GeV)$								&	$0.701\pm0.002\pm0.005$	\\	
$WZ\rightarrow f \bar{f'} l^{+}l^{-}$						&	$0.794\pm0.038\pm0.079$	&	$t \bar{t} \rightarrow q \bar{q}' b + e^+ \nu \bar{b}$			&	$0.709\pm0.004\pm0.041$	\\	
$Z/\gamma^{*}\rightarrow \mu^{+} \mu^{-}\,^{(m_{ll}>500\,GeV)}$\,&	$0.915\pm0.011\pm0.076$	&	$WZ\rightarrow l \nu f \bar{f}$					&	$0.655\pm0.012\pm0.075$	\\	
\hline
\end{tabular}
\label{tab:ExoticModelOverview}
}
\end{table}

\section{\label{Sec:Conclusion}Conclusion}

In this work, the model dependence of reinterpreting measured fiducial SM cross sections as a limit on BSM processes has been quantified for the first time using more than twenty SM and BSM processes. BSM models ranging from supersymmetric scenarios, to leptoquarks, to the impact of selected 6-dimensional effective field theory operators were considered in more than ten measurement fiducial volumes. The samples were generated with the \textsc{MadGraph} and \textsc{Pythia8} event generators, while the detector simulation was approximated with the \textsc{Delphes}-framework. 

The first, nearly trivial, however important conclusion is that the model-dependence can be significant when the number of final state objects differs between the SM process measured and the BSM process considered for reinterpretation. Concretely, differences were found between processes of two-lepton and three-lepton final states for a signal selection that requires exactly two leptons. Secondly, the model dependence is expected to be large when the signal selection cuts into any tails of observables with a limited resolution such as the reconstructed missing transverse energy of the event. Differences in the detector response corrections factors for different processes by up to 20\% have been observed. While special cases can certainly be constructed, where even larger differences are observed, in general the model dependence of fiducial cross-sections is not so large that a reinterpretation effort is not possible. An additional 20\% uncertainty might be therefore a first educated guess to cover model dependencies when interpreting fiducial cross-section measurements of Standard Model processes in view of new physics signatures. However, depending on the required precision, it might be important to correctly model the detector response for the BSM model under study and compare it to the SM process which is thought to be reinterpreted. Given that fast simulations typically do not describe tails of distributions well, it might be even required to use full simulations for reinterpretation of SM cross-section measurements when highest precision is required. 

\section*{Acknowledgements}

We would like to thank our collegues, in particular A. Buckley, J. Butterworth, T. Eifert, C. Gutschow, B. Malaescu, K. Monig and I. Vivarelli, for the useful discussions on this topic in the past years. The authors, M.S., A.S, K.M. would like to thank in addition, the Volkswagen Foundation for the support of this work.

\bibliographystyle{unsrt}
\bibliography{main}

\begin{thebibliography}{10}

\bibitem{Sjostrand:2007gs}
Torbjorn Sjostrand, Stephen Mrenna, and Peter~Z. Skands.
\newblock {A Brief Introduction to PYTHIA 8.1}.
\newblock {\em Comput. Phys. Commun.}, 178:852--867, 2008.

\bibitem{Gleisberg:2008ta}
T.~Gleisberg, Stefan. Hoeche, F.~Krauss, M.~Schonherr, S.~Schumann, F.~Siegert,
  and J.~Winter.
\newblock {Event generation with SHERPA 1.1}.
\newblock {\em JHEP}, 02:007, 2009.

\bibitem{Bahr:2008pv}
M.~Bahr et~al.
\newblock {Herwig++ Physics and Manual}.
\newblock {\em Eur. Phys. J.}, C58:639--707, 2008.

\bibitem{Alwall:2011uj}
Johan Alwall, Michel Herquet, Fabio Maltoni, Olivier Mattelaer, and Tim
  Stelzer.
\newblock {MadGraph 5 : Going Beyond}.
\newblock {\em JHEP}, 06:128, 2011.

\bibitem{Aaboud:2016zpd}
Morad Aaboud et~al.
\newblock {Measurements of top-quark pair to $Z$-boson cross-section ratios at
  $\sqrt s = 13, 8, 7$ TeV with the ATLAS detector}.
\newblock {\em JHEP}, 02:117, 2017.

\bibitem{CMS:2015ois}
CMS Collaboration.
\newblock {Measurement of inclusive W and Z boson production cross sections in
  pp collisions at sqrt(s)=13 TeV}.
\newblock {\em CMS-PAS-SMP-15-004}, 2015.

\bibitem{Dulat:2015mca}
Sayipjamal Dulat, Tie-Jiun Hou, Jun Gao, Marco Guzzi, Joey Huston, Pavel
  Nadolsky, Jon Pumplin, Carl Schmidt, Daniel Stump, and C.~P. Yuan.
\newblock {New parton distribution functions from a global analysis of quantum
  chromodynamics}.
\newblock {\em Phys. Rev.}, D93(3):033006, 2016.

\bibitem{Ball:2017nwa}
Richard~D. Ball et~al.
\newblock {Parton distributions from high-precision collider data}.
\newblock {\em Eur. Phys. J. C}, 77(10):663, 2017.

\bibitem{Ellis:1981tv}
John~R. Ellis, Dimitri~V. Nanopoulos, and Serge Rudaz.
\newblock {GUTs 3: SUSY GUTs 2}.
\newblock {\em Nucl. Phys.}, B202:43--62, 1982.

\bibitem{deBoer:1994dg}
W.~de~Boer.
\newblock {Grand unified theories and supersymmetry in particle physics and
  cosmology}.
\newblock {\em Prog. Part. Nucl. Phys.}, 33:201--302, 1994.

\bibitem{Schrempp:1984nj}
Barbara Schrempp and Fridger Schrempp.
\newblock {LIGHT LEPTOQUARKS}.
\newblock {\em Phys. Lett.}, 153B:101--107, 1985.

\bibitem{Pati:1974yy}
Jogesh~C. Pati and Abdus Salam.
\newblock {Lepton Number as the Fourth Color}.
\newblock {\em Phys. Rev.}, D10:275--289, 1974.
\newblock [Erratum: Phys. Rev.D11,703(1975)].

\bibitem{Dimopoulos:1979es}
Savas Dimopoulos and Leonard Susskind.
\newblock {Mass Without Scalars}.
\newblock {\em Nucl. Phys.}, B155:237--252, 1979.
\newblock [2,930(1979)].

\bibitem{Dorsner:2016wpm}
I.~Doršner, S.~Fajfer, A.~Greljo, J.~F. Kamenik, and N.~Košnik.
\newblock {Physics of leptoquarks in precision experiments and at particle
  colliders}.
\newblock {\em Phys. Rept.}, 641:1--68, 2016.

\bibitem{Djouadi:2012ae}
Abdelhak Djouadi and Alexander Lenz.
\newblock {Sealing the fate of a fourth generation of fermions}.
\newblock {\em Phys. Lett.}, B715:310--314, 2012.

\bibitem{Kribs:2007nz}
Graham~D. Kribs, Tilman Plehn, Michael Spannowsky, and Timothy M.~P. Tait.
\newblock {Four generations and Higgs physics}.
\newblock {\em Phys. Rev.}, D76:075016, 2007.

\bibitem{Frampton:1999xi}
Paul~H. Frampton, P.~Q. Hung, and Marc Sher.
\newblock {Quarks and leptons beyond the third generation}.
\newblock {\em Phys. Rept.}, 330:263, 2000.

\bibitem{Martin:2009bg}
Stephen~P. Martin.
\newblock {Extra vector-like matter and the lightest Higgs scalar boson mass in
  low-energy supersymmetry}.
\newblock {\em Phys. Rev.}, D81:035004, 2010.

\bibitem{ArkaniHamed:1998rs}
Nima Arkani-Hamed, Savas Dimopoulos, and G.~R. Dvali.
\newblock {The Hierarchy problem and new dimensions at a millimeter}.
\newblock {\em Phys. Lett.}, B429:263--272, 1998.

\bibitem{Randall:1999ee}
Lisa Randall and Raman Sundrum.
\newblock {A Large mass hierarchy from a small extra dimension}.
\newblock {\em Phys. Rev. Lett.}, 83:3370--3373, 1999.

\bibitem{Altarelli:1989ff}
Guido Altarelli, B.~Mele, and M.~Ruiz-Altaba.
\newblock {Searching for New Heavy Vector Bosons in $p \bar{p}$ Colliders}.
\newblock {\em Z. Phys.}, C45:109, 1989.
\newblock [Erratum: Z. Phys.C47,676(1990)].

\bibitem{London:1986dk}
David London and Jonathan~L. Rosner.
\newblock {Extra Gauge Bosons in E(6)}.
\newblock {\em Phys. Rev.}, D34:1530, 1986.

\bibitem{delAguila:2010mx}
F.~del Aguila, J.~de~Blas, and M.~Perez-Victoria.
\newblock {Electroweak Limits on General New Vector Bosons}.
\newblock {\em JHEP}, 09:033, 2010.

\bibitem{Aad:2014wea}
Georges Aad et~al.
\newblock {Search for squarks and gluinos with the ATLAS detector in final
  states with jets and missing transverse momentum using $\sqrt{s}=8$ TeV
  proton--proton collision data}.
\newblock {\em JHEP}, 09:176, 2014.

\bibitem{Chatrchyan:2013iqa}
Serguei Chatrchyan et~al.
\newblock {Search for supersymmetry in pp collisions at $\sqrt{s}$=8 TeV in
  events with a single lepton, large jet multiplicity, and multiple b jets}.
\newblock {\em Phys. Lett.}, B733:328--353, 2014.

\bibitem{Fairbairn:2006gg}
M.~Fairbairn, A.~C. Kraan, D.~A. Milstead, T.~Sjostrand, Peter~Z. Skands, and
  T.~Sloan.
\newblock {Stable massive particles at colliders}.
\newblock {\em Phys. Rept.}, 438:1--63, 2007.

\bibitem{Djouadi:1998di}
A.~Djouadi et~al.
\newblock {The Minimal supersymmetric standard model: Group summary report}.
\newblock In {\em {GDR (Groupement De Recherche) - Supersymetrie Montpellier,
  France, April 15-17, 1998}}, 1998.

\bibitem{Aitchison:2005cf}
Ian J.~R. Aitchison.
\newblock {Supersymmetry and the MSSM: An Elementary introduction}.
\newblock 2005.

\bibitem{Burgess:2007pt}
C.~P. Burgess.
\newblock {Introduction to Effective Field Theory}.
\newblock {\em Ann. Rev. Nucl. Part. Sci.}, 57:329--362, 2007.

\bibitem{Brivio:2017vri}
Ilaria Brivio and Michael Trott.
\newblock {The Standard Model as an Effective Field Theory}.
\newblock {\em Phys. Rept.}, 793:1--98, 2019.

\bibitem{Berthier:2015gja}
Laure Berthier and Michael Trott.
\newblock {Consistent constraints on the Standard Model Effective Field
  Theory}.
\newblock {\em JHEP}, 02:069, 2016.

\bibitem{deFavereau:2013fsa}
J.~de~Favereau, C.~Delaere, P.~Demin, A.~Giammanco, V.~Lemaître, A.~Mertens,
  and M.~Selvaggi.
\newblock {DELPHES 3, A modular framework for fast simulation of a generic
  collider experiment}.
\newblock {\em JHEP}, 02:057, 2014.

\bibitem{CERN:OpenData}
CERN.
\newblock Cern open data portal.
\newblock \url{http://opendata.cern.ch}.
\newblock (2019).

\bibitem{Aaboud:2016btc}
Morad Aaboud et~al.
\newblock {Precision measurement and interpretation of inclusive $W^+$ , $W^-$
  and $Z/\gamma ^*$ production cross sections with the ATLAS detector}.
\newblock {\em Eur. Phys. J.}, C77(6):367, 2017.

\bibitem{Aad:2010ey}
Georges Aad et~al.
\newblock {Measurement of the top quark-pair production cross section with
  ATLAS in pp collisions at $\sqrt{s}=7$ TeV}.
\newblock {\em Eur. Phys. J.}, C71:1577, 2011.

\bibitem{Aad:2012qf}
Georges Aad et~al.
\newblock {Measurement of the top quark pair production cross-section with
  ATLAS in the single lepton channel}.
\newblock {\em Phys. Lett.}, B711:244--263, 2012.

\bibitem{Cacciari:2008gp}
Matteo Cacciari, Gavin~P. Salam, and Gregory Soyez.
\newblock {The anti-$k_t$ jet clustering algorithm}.
\newblock {\em JHEP}, 04:063, 2008.

\bibitem{Aad:2019fac}
Georges Aad et~al.
\newblock {Search for high-mass dilepton resonances using 139 fb$^{-1}$ of $pp$
  collision data collected at $\sqrt{s}=$13 TeV with the ATLAS detector}.
\newblock {\em Phys. Lett.}, B796:68--87, 2019.

\end{thebibliography}

\clearpage
\appendix
\include{variables}

\end{document}